\documentclass[aps,prb,twocolumn,superscriptaddress,floatfix,amsmath,amssymb]{revtex4-2}
\usepackage{multirow}
\usepackage{graphicx}
\usepackage{bm}
\usepackage{dcolumn}
\usepackage{picture}
\usepackage[colorlinks=true,linkcolor=blue,urlcolor=blue,citecolor=blue]{hyperref}
\usepackage{natbib}
\usepackage{amsmath}
\usepackage{times}
\usepackage{color}
\usepackage{xcolor}
\usepackage{soul}
\usepackage{amssymb}

\begin{document}

\newcommand{\BNIO}{Ba$_4$NbIr$_3$O$_{12}$}

\title{Gapless dynamic magnetic ground state in the charge-gapped trimer iridate Ba$_4$NbIr$_3$O$_{12}$ }
\author{Abhisek Bandyopadhyay}
\email[email:]{abhisek.ban2011@gmail.com / abhisek.bandyopadhyay@stfc.ac.uk}
\affiliation{ISIS Neutron and Muon Source, STFC, Rutherford Appleton Laboratory, Chilton, Didcot, Oxon OX11 0QX, United Kingdom}

\author{S. Lee}
\affiliation{Center for Artificial Low Dimensional Electronic Systems, Institute for Basic Science, Pohang 37673, Republic of Korea}

\author{D. T. Adroja}
\email[email:]{devashibhai.adroja@stfc.ac.uk}
\affiliation{ISIS Neutron and Muon Source, STFC, Rutherford Appleton Laboratory, Chilton, Didcot, Oxon OX11 0QX, United Kingdom}
\affiliation{Highly Correlated Matter Research Group, Physics Department, University of Johannesburg, Auckland Park 2006, South Africa}

\author{M. R. Lees}
\affiliation{Department of Physics, University of Warwick, Coventry CV4 7AL, United Kingdom}

\author{G. B. G. Stenning}
\affiliation{ISIS Neutron and Muon Source, STFC, Rutherford Appleton Laboratory, Chilton, Didcot, Oxon OX11 0QX, United Kingdom}

\author{P. Aich}
\affiliation{Dipartimento di Scienze, Universitá Roma Tre, Via della Vasca Navale, 84 I-00146 Roma, Italy}
\affiliation{LASR3 Surface Analysis Laboratory, Universitá Roma Tre, Via della Vasca Navale, 84 I-00146 Roma, Italy}

\author{Luca Tortora}
\affiliation{Dipartimento di Scienze, Universitá Roma Tre, Via della Vasca Navale, 84 I-00146 Roma, Italy}
\affiliation{LASR3 Surface Analysis Laboratory, Universitá Roma Tre, Via della Vasca Navale, 84 I-00146 Roma, Italy}

\author{C. Meneghini}
\affiliation{Dipartimento di Scienze, Universitá Roma Tre, Via della Vasca Navale, 84 I-00146 Roma, Italy}

\author{G. Cibin}
\affiliation{Diamond Light Source Ltd., Diamond House, Harwell Science and Innovation Campus, Didcot, Oxfordshire OX11 0DE, UK}

\author{Adam  Berlie}
\affiliation{ISIS Neutron and Muon Source, STFC, Rutherford Appleton Laboratory, Chilton, Didcot, Oxon OX11 0QX, United Kingdom}

\author{R. A. Saha}
\affiliation{cMACS, Department of Microbial and Molecular Systems, KU Leuven, Celestijnenlaan 200F, Heverlee, 3001 Belgium}

\author{D. Takegami} 
\affiliation{Max Planck Institute for Chemical Physics of Solids, Nöthnitzer Straße 40, 01187 Dresden, Germany}
\affiliation{Department of Applied Physics, Waseda University, Shinjuku, Tokyo 169-8555, Japan}

\author{A. Meléndez-Sans} 
\affiliation{Max Planck Institute for Chemical Physics of Solids, Nöthnitzer Straße 40, 01187 Dresden, Germany}

\author{G. Poelchen} 
\affiliation{Max Planck Institute for Chemical Physics of Solids, Nöthnitzer Straße 40, 01187 Dresden, Germany}

\author{M. Yoshimura} 
\affiliation{National Synchrotron Radiation Research Center,
101 Hsin-Ann Road, Hsinchu 30076, Taiwan, R.O.C}

\author{K. D. Tsuei} 
\affiliation{National Synchrotron Radiation Research Center,
101 Hsin-Ann Road, Hsinchu 30076, Taiwan, R.O.C}

\author{Z. Hu} 
\affiliation{Max Planck Institute for Chemical Physics of Solids, Nöthnitzer Straße 40, 01187 Dresden, Germany}




\author{Ting-Shan Chan} 
\affiliation{National Synchrotron Radiation Research Center,
101 Hsin-Ann Road, Hsinchu 30076, Taiwan, R.O.C}
\author{S. Chattopadhyay}
\affiliation{UGC-DAE Consortium for Scientific Research, Mumbai Centre 247C, 2nd Floor, Common Facility Building
BARC Campus, Trombay, Mumbai-400085, INDIA}

\author{G. S. Thakur}
\affiliation{Department of Chemical Sciences, Indian Institute of Science Education and Research, Berhampur, Odisha, 760003, India}

\author{Kwang-Yong Choi}
\affiliation{Department of Physics, Sungkyunkwan University, Suwon 16419, Republic of Korea}

\date{\today}

\begin{abstract}
We present an experimental investigation of the magnetic ground state in Ba$_4$NbIr$_3$O$_{12}$, a fractional valent trimer iridate. X-ray absorption and photoemission spectroscopy show that the Ir valence lies between 3+ and 4+ while Nb is pentavalent. Combined dc/ac magnetization, specific heat, and muon spin rotation/relaxation ($\mu$SR) measurements reveal no magnetic phase transition down to 0.05~K. Despite a significant Weiss temperature ($\Theta_{\mathrm{W}} \sim -15$ to $-25$~K) indicating antiferromagnetic correlations, a quantum spin-liquid (QSL) phase emerges and persists down to 0.1~K. This state likely arises from geometric frustration in the edge-sharing equilateral triangle Ir network. Our $\mu$SR analysis reveals a two-component depolarization, arising from the coexistence of rapidly (90\%) and slowly (10\%) fluctuating Ir moments. Powder x-ray diffraction and Ir-L$_3$edge x-ray absorption fine structure spectroscopy identify ~8-10\% Nb/Ir site-exchange, reducing frustration within part of the Ir network, and likely leading to the faster muon spin relaxation, while the structurally ordered Ir ions remain highly geometrically frustrated, giving rise to the rapidly spin-fluctuating QSL ground state. At low temperatures, the magnetic specific heat varies as $\gamma T + \alpha T^2$, indicating gapless spinon excitations, and possible Dirac QSL features with linear spinon dispersion, respectively.
\end{abstract}

\maketitle

\section{Introduction}
The combined influence of spin-orbit coupling (SOC), electron correlations, geometric frustration, and competing magnetic superexchange interactions in the heavier 4$d$ and 5$d$ transition metal oxides leads to a diverse spectrum of novel magnetic and electronic properties, including quantum spin liquid (QSL) ground states~\cite{Kitaev1,Kim_prl,Hermanns_review,Kim_science,Haverkort_prl,Mizokawa_prl,Singh_prb,Chun_nature,okamotoprl2007,Takayama_jpsj,Perryprl2001,Cao_njp2004,Plumb_prb,Li_nature}. A QSL exhibits quantum fluctuations within its strongly correlated electron spins, and no magnetic order, even at temperatures approaching absolute zero. Such spin-liquid phases may host features such as fractionalized low-energy quasiparticle excitations (e.g. Majorana fermions, spinons, and emergent gauge flux)~\cite{Li_nature,Hao_qsl_musr,Balentsnature,Savary_review}
as well as hold potential for new technologies, including topological quantum computation~\cite{Kitaev1,Hao_qsl_musr,Balentsnature,Savary_review,Anderson,Kitaev2}. 



While, the 6$H$-hexagonal perovskite oxides Ba$_3$$M$$M^{\prime}_2$O$_9$ with $M^{\prime}_2$O$_9$ dimers ($M^{\prime}$ = 4$d$ or 5$d$ transition metal cation; $M$ = any mono, di-, tri-, or tetravalent nonmagnetic cation), have been extensively studied for novel quantum magnetic phases~\cite{Nagprl2016,Nguyen2021,Nagprb6H,Khanprb2019,Deyprb2014,Deyprb2017,Deyprb2013,Dey_epjb,Kumarprb2016,Garg_jpcm,Deyprb2012,Terasaki2017}, surprisingly, the trimer-based 12$L$-hexagonal quadrupole perovskite counterparts of general formula Ba$_4$$M$$M^{\prime}_3$O$_{12}$ ($M = $~rare earth; $M^{\prime} = $~heavier transition metal Ir, Rh, Ru, ...) have received more limited attention, despite their potential to harbor novel quantum magnetism including QSL phases. Previous studies primarily focused on their structure and bulk magnetic properties~\cite{Shimoda1,Shimoda2}. Recently, Nguyen {\it et al.} investigated three new systems, Ba$_4$Nb$M^{\prime}_3$O$_{12}$ ($M^{\prime}$ = Ir, Rh, Ru) as candidate QSL materials, highlighting the intriguing fractional valence of $M^{\prime}$ ions without charge ordering~\cite{Nguyenprm2018,Nguyenprm2019}. {\bf Thakur {\it et al.} used a combined magnetic and calorimetric study down to 2 K on off-stoichiometry Ba$_4$Nb$_{0.8}$Ir$_{3.2}$O$_{12}$ single crystals to reveal the possibility of a spin liquid state~\cite{Thakur}}. However, detailed microscopic magnetic investigations using local probes such as muon-spin-rotation/relaxation ($\mu$SR) or nuclear magnetic resonance were absent, leaving the true nature of the magnetic ground states and spin dynamics in these materials unclear. Very recently, we have investigated Ba$_4$NbRh$_3$O$_{12}$ using $\mu$SR, dc, ac magnetic susceptibilities, and heat capacity down to 0.05~K, revealing a QSL ground state~\cite{AbhisekarXiv}.

Here, we present comprehensive experimental data on a polycrystalline sample of Ba$_4$NbIr$_3$O$_{12}$ (BNIO) using a diverse range of techniques including powder x-ray diffraction (XRD), X-ray absorption (XAS) and high-resolution hard x-ray photoemission spectroscopies (HAXPES), electrical resistivity, dc and ac magnetic susceptibilities, specific heat, and muon spin rotation/relaxation ($\mu$SR). The combined XRD, XAS and chemical analysis using energy dispersive x-ray spectroscopy (EDX) revealed a single phase without stoichimetric defects. A fractional anti-site disorder of approximately 8-10\% between Nb and Ir atoms is found via XRD and XAS. While, x-ray absorption near-edge structure (XANES) and HAXPES data 
confirm the 
charge neutrality in this system via ensuring the cation valences to the desired numbers, a quantitative analysis of Ir-$L_2$ and $L_3$ edge XANES (white-line area) establishes 
a moderate effective SOC strength for Ir. 
The electrical resistivity together with valence band x-ray photoelectron spectroscopy (VBXPS) and {\it ab-initio} electronic structure calculations suggest a charge-gapped insulating electronic ground state. Despite significant antiferromagnetic interactions (Weiss temperature $\Theta_{\mathrm{W}} \sim -15$ to $-25$~K) among the sizeable local Ir moments ($\mu_{\mathrm{eff}}\approx$ 0.3 $\mu_{\mathrm{B}}$/Ir), our combined bulk dc and ac magnetic susceptibilities, specific heat, and $\mu$SR measurements indicate the absence of long-range magnetic order and/or a frozen magnetic ground state down to the lowest measurement temperature of 0.05~K. Instead, the system exhibits persistent spin dynamics as evidenced from our longitudinal-field (LF) $\mu$SR study at 0.1~K. This gives rise to a large frustration index, $f = |\frac{\Theta_{\mathrm{W}}}{T_{\mathrm{min}}}| \approx 400$ (where $T_{\mathrm{min}}$ is the lowest measured temperature down to which no magnetic ordering is observed), originating from the edge-sharing equilateral triangular network of Ir, which stabilizes a highly entangled dynamic QSL ground state. Furthermore, a remarkable universal scaling relation of the $\chi_{\mathrm{dc}}(T, H)$ and $M(H, T)$ in $T/H$ and $H/T$, respectively,
with the scaling exponent $\alpha_{\mathrm{s}} = 0.44 \approx \alpha_{\mathrm{m}} = 0.42$, is in agreement with highly frustrated quantum magnets \cite{cu2iro3prl,h3liir2o6prb,Kimchinature2018,Bahramiprl2019}.

Zero-field (ZF) and longitudinal field (LF) $\mu$SR measurements reveal the coexistence of a two-component muon depolarization  between 6~ and 0.1~K: $\sim10\%$ of slowly fluctuating Ir moments due to Nb/Ir anti-site-disorder on top of a strongly fluctuating QSL ground state arising from the structurally ordered Ir atoms ($\sim90\%$). The fluctuating (persistence of spin-dynamics) ground state is supported by the large retained magnetic entropy ($\sim 90\%$) in this material. The magnetic specific heat displays a clear $\gamma T + \alpha T^2$ dependence between 0.05 and 4~K. The linear contribution is related to the gapless nature of spinon excitations, while the presence of a finite $T^2$ term, in the absence of any magnetic order, could be the result of novel Dirac QSL phenomenology with a linear dispersion.
\section{Experimental techniques}
Polycrystalline \BNIO was prepared using a standard solid-state reaction technique. Stoichiometric ratios of the starting materials, Ba$_2$O$_3$, Nb$_2$O$_5$, and IrO$_2$ (Sigma Aldrich, 99.95\%, 99.9\% and 99.8\%, respectively) were mixed and finely ground in an agate mortar, then heated to 800~$^{\circ}$C for 24~hours in air. The calcined powders were reground, pressed into pellets, and sintered at 1100~$^{\circ}$C for 48 hours
with several intermediate grindings. The phase purity and crystal structure of the samples were determined through XRD using a Bruker D8 Advance diffractometer with Cu-$K_{\alpha}$ radiation. A structural refinement was performed by the Rietveld technique using FULLPROF~\cite{Fullprof}. The chemical homogeneity and cation stoichiometry of this sample were verified in a ZEISS Supra 55VP scanning electron microscope with an Oxford Instruments energy-dispersive X-ray (EDX) spectrometer. The electrical resistivity was measured as a function of temperature ($\rho$ versus $T$ ) in a Quantum Design Physical Property Measurement System (QD-PPMS) using a standard four-probe method over the temperature range of 80 to 400~K with a measurement current of 10 $\mu$A. Core-level and valence band HAXPES measurements were carried out at room temperature at the HAXPES end station of the Max Planck–National Synchrotron Radiation Research Center (NSRRC) at the Taiwan undulator beamline BL12XU at SPring-8, Japan. Measurements were made using linearly polarized light with a photon energy of 6.5~keV and an energy resolution of 280~meV, using a MB Scientific A-1 HE analyzer mounted parallel to the polarization direction~\cite{Takegami2019}. To corroborate the HAXPES data, we computed the partial density of states (PDOS) by means of fully relativistic density functional theory (DFT) calculations within the local density approximation (LDA) using the full-potential local-orbital (FPLO) code~\cite{Koepernik1999}.
Ir-L$_2$ and $L_3$ edge XAS measurements were carried out at the B-18 beamline of the Diamond Light Source, UK, in standard transmission geometry at low temperature (10~K).
The Nb-$L_3$ XAS data were collected at the BL16A beamline of NSRRC in Taiwan. The near edge (XANES) and extended (EXAFS) regions of the XAS spectra were treated following the standard procedures for background subtraction and normalization, and analyzed using freely available packages (Demeter~\cite{Newville,Ravel}, Estra-Fitexa~\cite{Fitexa}, Fityk~\cite{Fityk}). A multishell EXAFS data refinement procedure was carried out as described in Refs.~\cite{Middeyprb2011,Middeyprb2016} to characterize the Ir local atomic structure. DC magnetic susceptibility, $\chi$, as a function of temperature was measured using a Quantum Design MPMS3 superconducting quantum interference device equipped with a vibrating sample magnetometer in the temperature range of 2-300~K and in applied magnetic fields $H$ of up to $\pm60$~kOe. The temperature-dependent specific heat in applied magnetic fields of up to 90~kOe was measured between 2 and 300~K in a QD-PPMS a using the $2\tau$ relaxation method. Specific heat, $C_p$, and ac magnetic susceptibility, $\chi_{\mathrm{ac}}$, data were also collected between 0.05 and 4~K using a QD Dynacool PPMS. Zero-field (ZF) and longitudinal-field (LF) muon spin rotation/relaxation ($\mu$SR) measurements were conducted using the MuSR spectrometer at the ISIS Neutron and Muon Source, UK. For $\mu$SR measurements, the powder sample of \BNIO\ was mounted on a 99.995\% silver holder using GE varnish to ensure good thermal contact. The sample was covered with a thin Ag-foil and loaded in a dilution refrigerator that was used to cool down the sample down to 0.1~K. The $\mu$SR asymmetry spectra were analyzed using the MANTID software package~\cite{Mantid}.

\section{Results and Discussion}
\subsection{Structural and chemical characterization from X-ray diffraction and EDX}
\begin{figure*}[tbh!]
\begin{center}
{\includegraphics[width=0.70\linewidth]{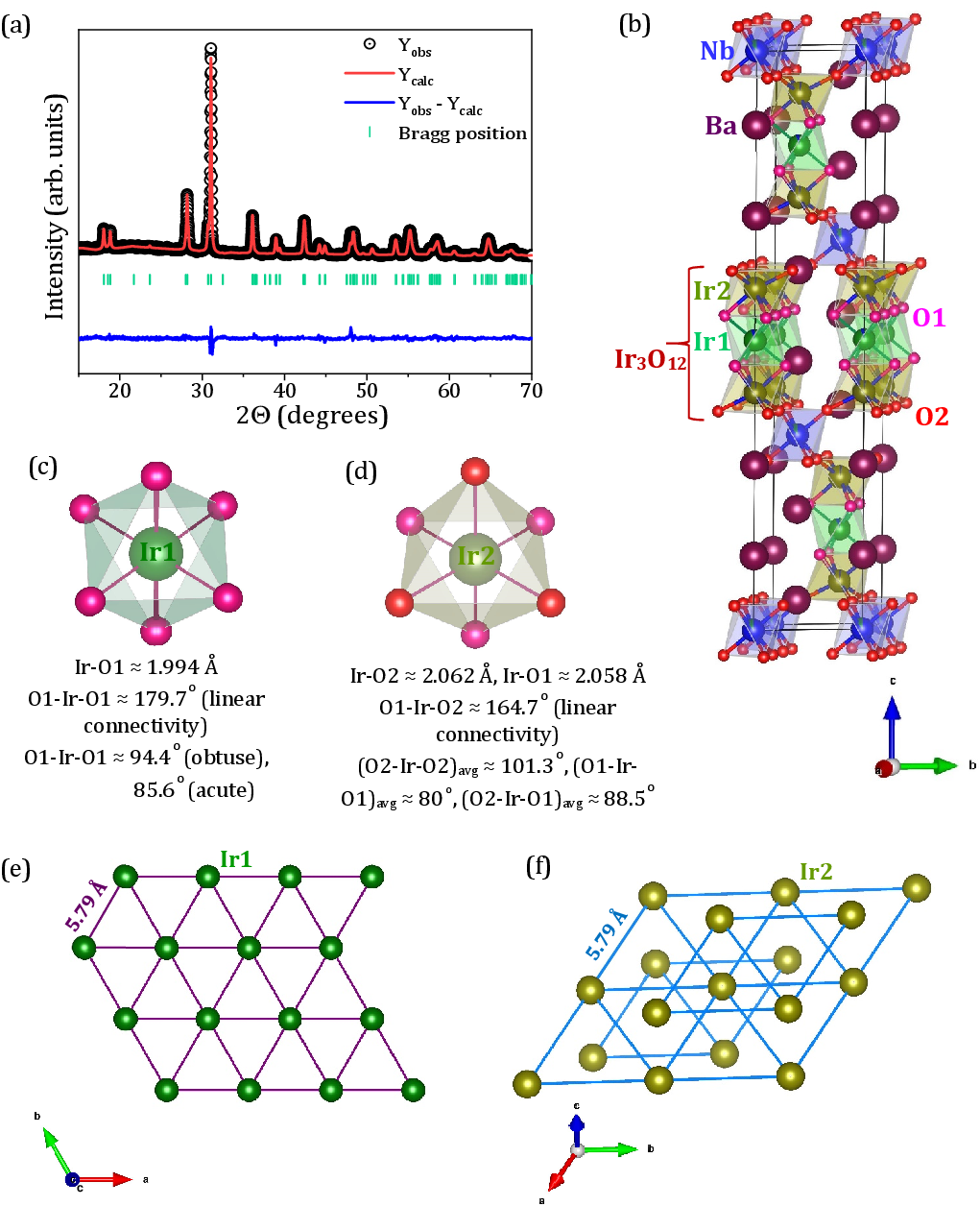}}
\caption{(a) Rietveld refined powder x-ray diffraction pattern for \BNIO~at 300~K, along with the difference (Y$_{\mathrm{obs}}$-Y$_{\mathrm{cal}}$, shown by the blue solid  line), and the allowed Bragg peaks (green vertical ticks). (b) Refined crystal structure, with the atoms shown in different colors and the Ir$_3$O$_{12}$ trimer units shaded in green. (c) and (d) Two distinct IrO$_6$ octahedra with corresponding Ir-O bond lengths and O-Ir-O bond angles. (e) and (f) Edge-sharing equilateral triangular network of Ir ions (shown separately for Ir1 and Ir2 with distinct colors and bonds).}
\label{FIG:Structure}
\end{center}
\end{figure*}

\begin{table}[tbh!]
\begin{center}
\caption{Structural parameters obtained from the Rietveld refinement of the powder x-ray diffraction data of \BNIO. Space group, $R\bar{3}m$, $a = b = 5.7590(4)$~{\AA}, $c = 28.6974(3)$~\AA, $\gamma = 120^{\circ}$, $V = 824.278(13)$~\AA$^3$; $R_{\mathrm{p}} = 12.1$, $R_{\mathrm{wp}}$ = 16.1, $R_{\mathrm{exp}} = 9.77$, $\chi^2 = 2.72$, $\chi^2R_{\mathrm{Bragg}}=2.93$. $B$ is the isotropic temperature factor. The occupancies for each of the Nb1-Ir1, Nb2-Ir2, and Nb3-Ir3 pairs are constrained to sum to one.}
\label{Table:refined}
\resizebox{8.6cm}{!}{
\begin{tabular}{ c  c  c  c  c  c  c }
\hline\hline Atom   & Site & Occupancy &  $x$ & $y$ & $z$ & $B$({\AA}$^2$) \\\hline
  Ba1 & 6c & 1.000  & 0 & 0       & 0.1283(3)    & 0.009(6) \\
  Ba2 & 6c & 1.000  & 0 & 0       & 0.2849(4)    & 0.009(3) \\
  Nb1 & 3a & 0.9019(4) & 0 & 0       & 0            & 0.012(3)\\
  Ir & 3a & 0.098(1) & 0 & 0       & 0            & 0.016(5)\\
  (site-disordered with Nb1) &  &  &  &  &  &  \\
  Ir1 & 3b & 0.9452(4) & 0 & 0       & 0.5          & 0.018(2)\\
  Nb & 3b & 0.0548(2) & 0 & 0       & 0.5          & 0.012(6)\\
  (site-disordered with Ir1) &  &  &  &  &  &  \\
  Ir2 & 6c & 0.9743(5) & 0 & 0       &  0.4098(1) & 0.018(5)\\
  Nb & 6c & 0.0257(1) & 0 & 0       & 0.4098(1)  & 0.012(5)\\
  (site-disordered with Ir2) &  &  &  &  &  &  \\
  O1 & 18h & 1.000  & 0.5024(2) & 0.5076(2)  & 0.1247(1) & 0.045(6)\\
  O2 & 18h & 1.000  & 0.5206(2) & 0.4834(2)  & 0.2893(2) & 0.046(3)\\
\hline\hline
\end{tabular}
}
\end{center}
\end{table}
Figure~\ref{FIG:Structure}(a) shows a Rietveld refinement of the powder x-ray diffraction pattern for polycrystalline \BNIO~at room temperature. The sample appears single phase (no additional unindexed reflections are visible in the residuals) with an $R\bar{3}m$ space group, in accord with an earlier report by Nguyen {\it et al.}~\cite{Nguyenprm2018}. The refined structural parameters along with the goodness-of-fit factors are summarized in Table~\ref{Table:refined}. The refinement improves statistically by allowing chemical disorder between the Nb and Ir sites, resulting in $\sim$8-10\% Ir/Nb occupancy switch. This is not surprising given the similar ionic radii R of six coordinated Nb$^{5+}$, (R$_{\mathrm{Nb}^{5+}}\sim 0.64$~\AA), and six coordinated Ir (R$_\mathrm{{Ir}^{3+}}\sim 0.68$ \AA\ and R$_{\mathrm{Ir}^{4+}}\sim 0.63$ \AA). For reference, a Rietveld refinement of the XRD pattern for perfect Nb/Ir site ordering is shown in Appendix~\ref{XRD data data}. The absence of any super-lattice peaks excludes the possibility of any long-range charge and/or chemical (Ir/Nb) ordering within the Ir-timer network.
Figure~\ref{FIG:Structure}(b) shows the refined crystal structure, consisting of three face-sharing IrO$_6$ octahedra forming an Ir$_3$O$_{12}$ trimer, which is coupled to its neighboring Ir$_3$O$_{12}$ trimer through corner-sharing nonmagnetic NbO$_6$ octahedra along the $c$ axis. Such a structural arrangement naturally suggests a stronger intra-trimer Ir-Ir magnetic exchange coupling (via Ir-Ir direct exchange and Ir-O-Ir superexchange pathways) compared to the inter-trimer Ir-Ir magnetic exchange interaction via longer Ir-O-(Nb)-O-Ir super-superexchange pathways. 

The IrO$_6$ octahedra are different for the two inequivalent Ir1 and Ir2 sites [see Figs.~\ref{FIG:Structure}(c) and \ref{FIG:Structure}(d)], providing different non-cubic crystal field effects mainly due to rotational (Ir-O bonds between opposite oxygen atoms are misaligned: O-Ir1/2-O bond angles $\ne 180^{\circ}$) and trigonal (Ir-O bonds between the other oxygen atoms are not perpendicular, O-Ir-O angles $\neq 90^{\circ}$) distortions around the respective Ir ions. However, the local magnetic environments around the individual Ir-sites would not be drastically different based on the fact that the two types of Ir-O (Ir2-O1 and Ir2-O2) bond distances in the Ir2O$_6$ octahedra  can be treated as six similar Ir2-O distances (while Ir1O$_6$ octahedra possess six Ir-O bond lengths of one type). This would not lead to significant variation in the lifting of the $t_{2g}$ orbital degeneracy corresponding to two distinct IrO$_6$-octahedral sites in this material. {\bf A reliable assignment of the Ir charges within the Ir$_3$O$_{12}$ trimers therefore cannot be made. However, implementing the bond valence sum (BVS) model at each of the Ir sites gives an rough estimate of the Ir-valence $V_i$ at site $i$. $V_i = \sum_j S_{ij}$ is a sum over the neighboring atoms $j$ and $S_{ij}$ is approximated by $S_{ij} = \exp{(R_0 - R_{ij})/B}$, where $R_{ij}$ is the interatomic distance between atoms $i$ and $j$, while $R_0$ and $B$ are constant bond valence parameters~\cite{Brown2009}. This gives Ir valences of 3.4(1) and 3.6(1) at the Ir1 and Ir2 sites, respectively.} Furthermore, both the Ir1 and Ir2 ions of the Ir$_3$O$_{12}$ trimer constitute a highly geometrically frustrated equilateral triangular network, as shown in Figs.~\ref{FIG:Structure}(e) and \ref{FIG:Structure}(f).

Energy-dispersive X-ray spectrometry confirms that to within the accuracy of the EDX technique, our \BNIO~sample is chemically homogeneous.

\subsection{Local structural study from extended x-ray absorption fine structure (EXAFS)}
\begin{table}[tbh]
\begin{center}
\caption{Local structure parameters extracted from the Ir-$L_3$ edge EXAFS data analysis. SS: single scattering; MS: multiple scattering; $x_{\mathrm{AS}}$: Fraction of Ir/Nb antisite disorder}
\label{Table:exafs}
\resizebox{7.5cm}{!}{
\begin{tabular}{ l  l  c  c  c }
\hline\hline 
 Neighbor shell & $N$    & $\sigma^2$             & $R$   &  $R_{\mathrm{XRD}}$ \\
                 &        & $(10^{-3}$ {\AA}$^2$) & (\AA) &  (\AA) \\
 \hline
  O             & 6       & 2.2(1)   & 2.034(2)   &  2.038  \\
  Ir/Nb         & \multirow{2}{*}{1.33} 
                          & \multirow{2}{*}{2.5(2)} 
                                     & \multirow{2}{*}{2.553(5)}  
                                                  &  \multirow{2}{*}{2.59}\\
\multicolumn{1}{r}{$x_{\mathrm{AS}}=0.09(3)$} &  &  &   &        \\
  Ba$_1$        & 4       & \multirow{2}{*}{4.1(3)}   & 3.49(2)  &   3.46     \\
  Ba$_2$        & 2.7     &          & 3.63(3)  &   3.63     \\
  Nb(SS+MS)     & 2       & 1.56(1)  & 3.98(5)  &   4.02  \\
\hline\hline
\end{tabular}}
\end{center}
\end{table}
For complex oxide materials, how the ions are locally arranged and connected  (i.e., local-order) plays a crucial role in shaping their magnetic and electrical properties. This local atomic structure can sometimes differ from the crystallographic structure indicated by XRD, which only accounts for structural features with global symmetry~\cite{Middeyprb2011,Middeyprb2016}.
Chemical selective and local order sensitive x-ray absorption fine structure (XAFS) spectroscopy provide reliable details about the local coordination geometry.
\begin{figure}[tbh!]
\begin{center}
{\includegraphics[width=\linewidth]{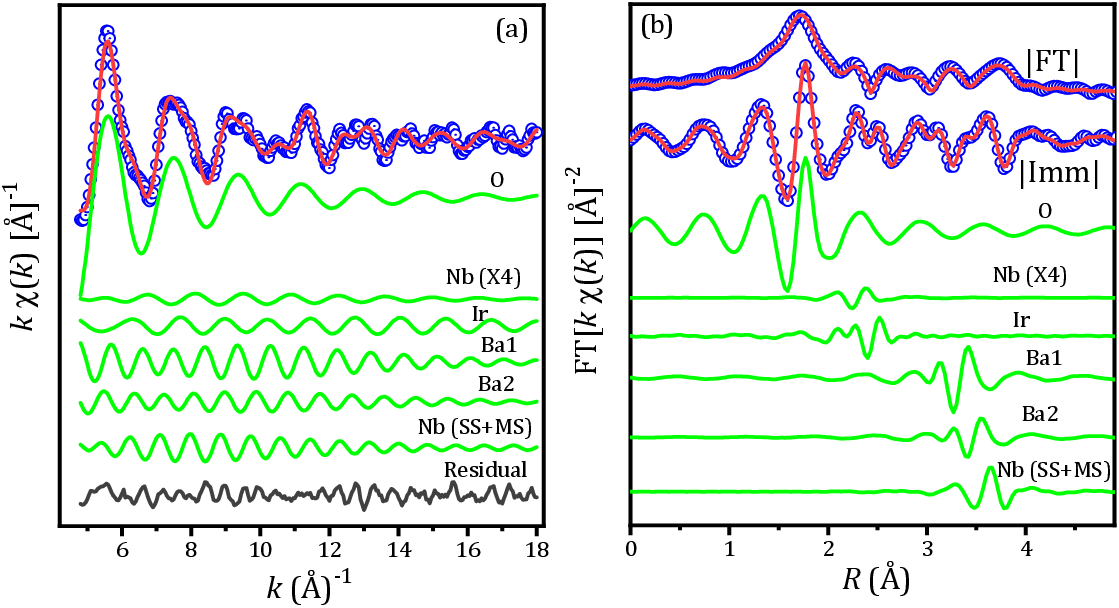}}
\caption {(a) Ir $L_3$-edge $k^2$ weighted experimental EXAFS data (open blue circles) along with the corresponding best fits (red solid lines) in the 3.5 to 19~{\AA}$^{-1}$ $k$ range. The individual single scattering (SS) and multiple scattering (MS) paths are shown in green solid lines along with the residuals as a black solid line. (b) Fourier transforms of the experimental data (blue open circles) and the theoretical fitted curve (red solid line). The real ($|$FT$|$) and imaginary ($|$Imm$|$) parts are shifted vertically for clarity.}
\label{FIG:exafs}
\end{center}
\end{figure}
Ir-$L_3$ edge XAFS data for \BNIO~are shown in Figs.~\ref{FIG:exafs} (a) and ~\ref{FIG:exafs}(b). 
Data refinement was performed using the crystallographic structure obtained from the XRD data analysis as a starting atomic cluster model around the average Ir site, using both the Artemis~\cite{Newville,Ravel} and Fitexa~\cite{Fitexa} programs, providing the fully consistent results summarized in Table~\ref{Table:exafs}. The photoelectron amplitude and phase functions were calculated using the FEFF program~\cite{FEFF}, while model curves were calculated using the standard EXAFS formula~\cite{EXAFS}. The multiplicity numbers were kept fixed to the crystallographic values. In order to account for the  Ir/Nb antisite disorder, a fraction ($x_{\mathrm{AS}}$) of Ir neighbors was substituted by Nb using the same Debye-Waller factor and distance for Ir and Nb contributions (Table~\ref{Table:exafs}). The refined value $x_{\mathrm{AS}} = 0.09\pm0.03$ is consistent with our XRD findings and the Ir-Ir/Nb distance (2.55~\AA) is in good agreement with the XRD result (2.59~\AA). The average 6.67 Ir-Ba neighbors are broadly distributed in the crystallographic model. They have been considered in two sub-shells of around 3.49 and 3.63~\AA, in good agreement with the crystallographic structure. Finally, we found the structural signal from Ir-O-Nb configurations relevant in the EXAFS data refinement. Here the multiple scattering (MS) signal is enhanced due to the almost collinear configuration and dominates the structural signal. 
The mean square relative displacement ($\sigma^2$) of the Ir-O coordination shell obtained from the EXAFS analysis ($2.2 \times 10^{-3}$~\AA$^2$) is much smaller than the variance of crystallographic Ir-O distribution ($\sigma^2_{\mathrm{XRD}} = 1.0 \times 10^{-2}$~\AA$^2$). This cannot be an effect of temperature, as we have also observed such a difference in the 250~K Ir-$L_3$ edge XAFS data analysis [not shown here], thus indicating that the Ir-O octahedra are quite regular and weakly distorted.

\subsection{Electronic characterization}
\subsubsection{\bf Ir $L_3$ edge XANES}
Given the stoichiometry of \BNIO, the oxidation states of the metal Ir and Nb cations should be 3.67+ and 5+, respectively, in order to maintain charge neutrality. Covalency-driven charge disproportionation and the resulting mixed valency of the transition metal ions prevail as the crucial factors which often modify the magnetic ground-state behavior in many of the complex magnetic oxides. 
\begin{figure*}[tbh!]
\begin{center}
{\includegraphics[width=0.80\linewidth]{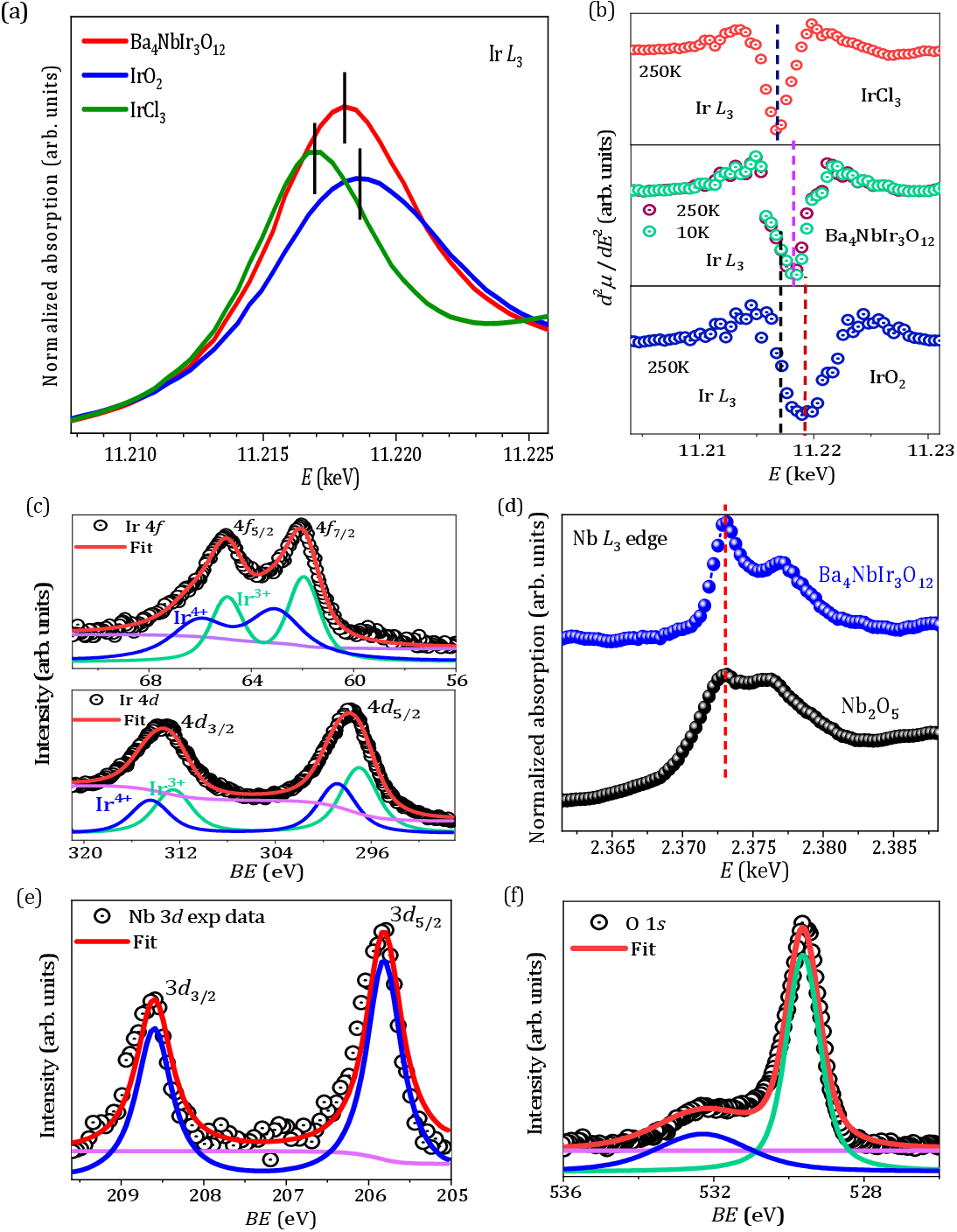}}
\caption{(a) XANES spectra at the Ir $L_3$ edge for \BNIO , IrCl$_3$, and IrO$_2$ at 250 K. 
(b) Second derivative curves of the normalized XANES spectra, indicating a white line feature. (c) Core-level Ir-4$f$ (upper panel) and 4$d$ (lower panel) HAXPES spectra (shaded black circles) as a function of binding energy (BE) along with fits (red solid lines). (d) Nb $L_3$-edge XAS spectrum of \BNIO~along with Nb$_2$O$_5$ for comparison; the curves are vertically shifted for clarity. (e) Nb-3$d$ core-level HAXPES spectra (open black circles) along with the fits (red solid line). (f) O-1$s$ core-level x-ray photoemission spectra (black open circles), along with theoretical fit (red solid line).}
\label{FIG:xanes}
\end{center}
\end{figure*}
In order to evaluate the average cation valences in \BNIO, we  combined Ir-$L_3$ near edge and Ir-4$f$ core-level HAXPES (at 300~K) analysis. The intense peak [white-line, (WL)] observed at the Ir $L_3$-edge in XANES, originates from 2$p \rightarrow 5d$ electronic transition, and its energy is directly related to the Ir-oxidation state~\cite{Iroxi1,Iroxi2}. Figure~\ref{FIG:xanes}(a) shows the Ir $L_3$ XANES region measured for \BNIO, and IrCl$_3$ and IrO$_2$ references. 
The high energy shift of the white-line feature (the rising edge, see in Fig.~\ref{FIG:xanes}(a)) reflects the raising of Ir oxidation state from Ir(3+)Cl3 to Ir(4+)O2, which is further highlighted by looking at the corresponding second derivative curves presented in Fig.~\ref{FIG:xanes}(b). Assuming a linear trend in the WL position as a function of the average valence state, the estimated Ir-valence is in agreement with the expected stoichiometric value, 3.67+. The iridium based materials with Ir-valence $\geq$ 4+ possess well-resolved doublet features in the respective WL spectra, corresponding to the Ir 2$p \rightarrow t_{2g}$ (lower-energy feature) and 2$p \rightarrow e_g$ (higher-energy peak) transitions, while for systems with Ir$^{3+}$, the lower-energy 2$p \rightarrow t_{2g}$ transition is absent due to completely filled $t_{2g}$ orbitals (Ir$^{3+}$:5$d^6$, i.e., six 5$d$ electrons in triply degenerate $t_{2g}$ orbitals), instead producing a single symmetric peak in the second derivative curve. Considering all this, the observed peak shape, asymmetry, energy positions, and the overall spectral structure of \BNIO~relative to those of the reference materials Ir$^{3+}$Cl$_3$ and Ir$^{4+}$O$_2$ [see Fig.~\ref{FIG:xanes}(b)] supports the idea that the Ir charge in \BNIO~lies between 3+ and 4+.
\subsubsection{\bf Ir 4$f$ core-level HAXPES}
Experimental XPS core-level spectra of Ir 4$f$ and 4$d$, are shown in the top and bottom panels, respectively, of Fig.~\ref{FIG:xanes}(c). In both the Ir 4$f$ and the 4$d$ core level spectra, the peaks display strongly asymmetric shapes that can be well reproduced by accounting for two spin-orbit doublets, as shown with the green and blue curves.  

If we associate the lower binding energy doublet (green doublet) to the Ir$^{3+}$ charge state, and the higher binding energy doublet (blue doublet) to the Ir$^{4+}$ charge state, we find that the fraction of the Ir$^{4+}$ charge with respect to the Ir$^{3+}$ species, estimated from the ratio of the respective integrated intensity [i.e., ratio of the area under the spin-orbit doublet of Ir$^{4+}$ (blue curve) to that of Ir$^{3+}$ (green curve) from both the Ir-4$f$ and 4$d$ core-level HAXPES spectra shown in Fig.~\ref{FIG:xanes}(c)], are close to $\sim$ 0.67, which is in good agreement with the average 3.67+ valence of Ir in \BNIO.

The spin-orbit energy separation for each of the fitted doublet components for both the 4$f$ and 4$d$ photoemission spectra and the ratio of the respective doublet peak intensities remain fixed at their theoretically desired values during the fits. {\bf However, such conventional fitting procedures and assignments of Ir4+ and Ir3+ charge states are only an approximation that might not be fully applicable in a strongly interacting case like \BNIO\ where the final states present mixed character.}

\subsubsection{\bf Combined Nb $L_3$ XAS and Nb 3$d$ core-level HAXPES}
The Nb-valence was checked using x-ray absorption spectroscopy at the Nb $L_3$ edge [Fig.~\ref{FIG:xanes}(d)] and core-level Nb-3$d$ HAXPES [Fig.~\ref{FIG:xanes}(e)]. As shown in Fig.~\ref{FIG:xanes}(d), the Nb-L$_3$ XAS spectrum of \BNIO~lies exactly at the same energy as that of Nb$_2$O$_5$, indicating a Nb$^{5+}$ valence state. The Nb 3$d$ core-level XPS spectrum, shown in Fig.~\ref{FIG:xanes}(e), was theoretically modeled using a single spin-orbit doublet featuring the 3$d_{5/2}$ (lower-BE at 205.75~eV) and 3$d_{3/2}$ (higher-BE at 208.66~eV) peaks. The energy positions of the peaks in the doublet along with their spin-orbit separation ($\sim 3$~eV), support a 5+ valence of Nb in this system.
\subsubsection{\bf O 1$s$ HAXPES}
In the O 1$s$ core-level HAXPES spectrum [see in Fig.~\ref{FIG:xanes}(f)], instead of a clean singlet, a pronounced higher-BE shoulder is evident, indicating mixed chemical environments. As shown, the O 1$s$ spectrum has been fit by considering two singlets. The contribution of the higher-BE singlet [blue curve in Fig.~\ref{FIG:xanes}(f)] is about $\sim 10\%$ of the lower-BE primary singlet [green curve in Fig.~\ref{FIG:xanes}(f)], which agrees well with the $\sim 8-10\%$ Nb/Ir site-exchange as obtained from both the XRD and EXAFS analysis. The higher-BE singlet in the O 1$s$ spectrum is therefore ascribed to the presence of Nb/Ir antisite disorder in \BNIO.
\subsubsection{\bf Valence band x-ray photoemission spectroscopy and theoretical study}
\begin{figure*}[tbh!]
\begin{center}
{\includegraphics[width=1.6\columnwidth]{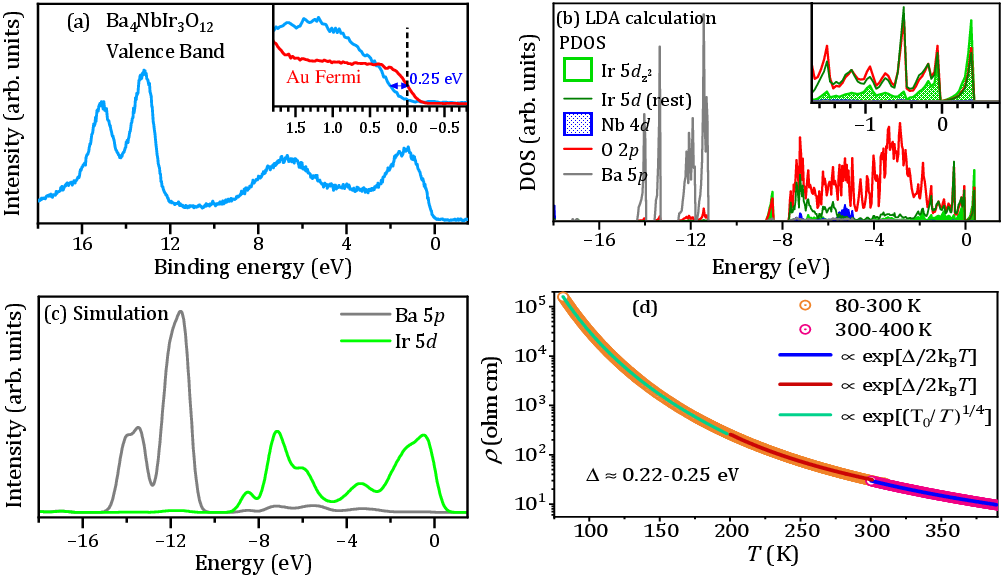}}
\caption{(a) Valence band photoemission spectrum for \BNIO. Inset: Expanded view of the same data close to the Fermi energy (solid cyan line) showing an energy gap. The valence band spectrum (red solid line) for Au metal was measured as a reference to obtain an accurate energy calibration. (b) Calculated partial density of states (PDOS) of the principal contributions in \BNIO. The Ir~$5d$ contributions are separated into projections along the $z$ axis (light green) and the rest (dark green). (c) Simulated HAXPES valence band spectrum, obtained by multiplying the PDOS of the relevant occupied states shown in the top panel by their respective photoionization cross sections at 6.5~keV photon energy, followed by a broadening to account for experimental conditions. (d) Temperature-dependence of the zero-field electrical resistivity between 80 and 400~K (open circles) along with the fits (solid lines) for different temperature regions.}
\label{FIG:VB}
\end{center}
\end{figure*}
In order to examine the nature of electronic ground state in this material, valence band x-ray photoemission spectroscopy data were collected and are displayed in Fig.~\ref{FIG:VB}(a). The absence of the density of states right at the Fermi level confirms the charge-gapped insulating behavior of this material. To better understand the experimentally collected VBXPS spectrum, we performed density functional theory calculations, and the results are shown in Fig.~\ref{FIG:VB}(b). The calculated partial density of states shows that the O~$2p$ bands are widely spread between 8 and 0~eV. The Ir~$5d$, PDOS has two main features at around 8 and 1~eV, corresponding to the bonding and antibonding from strong hybridization with the O~$2p$ orbitals~\cite{Takegamiprb2020}. However, the Ir~$5d_{z^2}$ displays a larger splitting than the other contributions, indicating a stronger bond along the trimer direction. The Nb~$4d$ PDOS is mainly located around 5~eV. While the Nb features are accompanied by O~$2p$ PDOS, indicating the presence of Nb-O hybridization, there is barely any overlap between the Ir~$5d$ and Nb~$4d$ features, indicating there is no significant interplay between the Ir and Nb. The Ba $5p$ semi-core level shows two deep peaks split by SOC, corresponding to the $5p_{3/2}$ and $5p_{1/2}$ contributions, as well as a small amount of weight in the PDOS in the valence region due to weak hybridization to the valence bands.  Our LDA calculations provide a gapped solution [Fig.~\ref{FIG:VB}(b)], indicating that the material is already a band insulator. 
Finally, we simulate a spectrum from the PDOS, by multiplying each of the PDOS by its photoionization cross section as well as the Fermi function, and then broadening the result [see Fig.~\ref{FIG:VB}(c)]. There is a reasonable match to the experiment. The spectral weight of the experimental valence band can largely be explained by the Ir $5d$ bonding-antibonding peaks, which dominate the spectra due to their large cross sections, mixed with some contributions from the Ba $5p$ \cite{Takegami2019}.

{\bf Appendix~\ref{Branching ratio (BR)} shows an expanded view of the calculated band structure of \BNIO. The band structure of the Ruddlesden-Popper phases Sr$_2$IrO$_4$ and  Ba$_2$IrO$_4$~\cite{Arita_prl,Agrestini_prb} and the double perovskite Ba$_2$YIrO$_6$~\cite{Dey_prb2016}, reveal broad t$_{2g}$ bands that are observed to cross one another. These materials are often discussed in the context of a $J_{\mathrm{eff}} = \frac{1}{2}$ Mott ground state~\cite{Kim_prl}. In contrast, more distinct, split, flatter bands are more characteristic of molecular orbitals, as seen in Nb$_3$Cl$_8$ \cite{Gao_prx2023}. The formation of molecular orbitals leads to the lifting of degeneracies and formation of discrete states according to their symmetries. The flattening is due to the localization of such states around the trimer/dimer. In \BNIO, we observe that the Ir $t_{2g}$ bands are split into distinct states with highly mixed character and these different states avoid crossing one another (e.g. the first 5 bands are split into 1-2-2 distinct states), suggesting the formation of ``molecular orbital''-like mixed orbitals within the Ir$_3$O$_{12}$ trimers. Together with the weak SOC in \BNIO\ (see Section~\ref{SOC} below) this suggests that a $J_{\mathrm{eff}} =  \frac{1}{2}$ picture may not be appropriate for \BNIO.} 

\subsubsection{\bf Temperature dependent electrical resistivity}
The temperature dependence of the electrical resistivity, $\rho(T)$, of \BNIO~is shown in Fig.~\ref{FIG:VB}(d). $\rho(T)$ increases with decreasing temperature from $\sim 10~\Omega$-cm at 400~K to about $10^5~\Omega$-cm at 80~K, above which the resistance exceeds the maximum limit of the instrument. $\rho(T)$ follows an Arrhenius activated behavior, with an energy gap of $\Delta \sim 0.22-0.25$~eV [consistent with the VBXPS measurement, shown in the inset of Fig.~\ref{FIG:VB}(a)] between 200 to 400~K, and then follows a 3-D variable-range-hopping mechanism from $\sim 200$~K down to the lowest measurement temperature of 80~K.


\subsection{\label{SOC}Estimate of the effective Ir-SOC from the Ir-XANES and the branching ratio}
\begin{figure}[tbh!]
\begin{center}

{\includegraphics[width=0.90\linewidth]{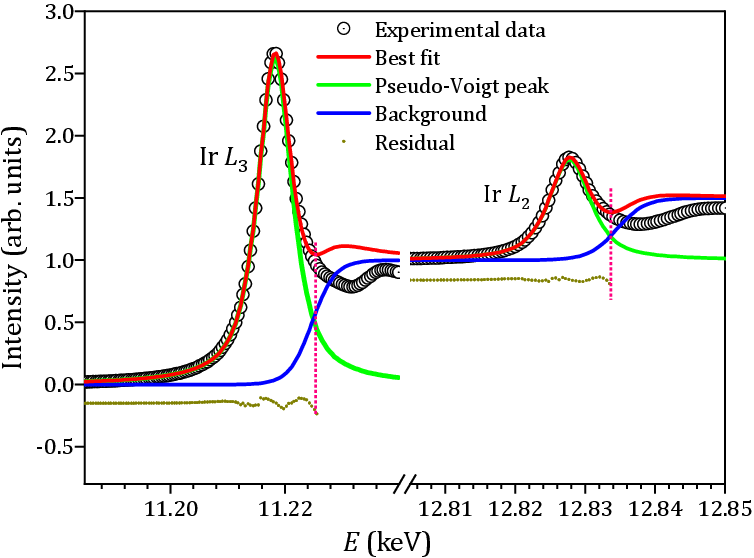}}
\caption{Ir-$L_3$ and $L_2$ edges XANES spectra from \BNIO, along with the theoretical fits to the $L_3$ and $L_2$ edge white-line intensities using a combination of pseudo-Voigt peak and a sigmoid background function.}
\label{FIG:BR}
\end{center}
\end{figure}

It is crucial to have a measure of the effective Ir-5$d$ SOC in \BNIO, as in the 5$d$ iridium oxides, atomic spin-orbit coupling effects can be masked by several factors (e.g. noncubic crystal distortions, electronic bandwidth, hopping, hybridization, superexchange, etc.)~\cite{Takegamiprb2020,snioownprb,Manjilprb2022,Liuprl2012,Calderprb2017,Nagprl2019}. The Ir-$L_2$ and $L_3$ edge white-line intensities in XANES have been carefully investigated [see Appendix~\ref{Branching ratio (BR)} for a detailed discussion of the branching ratio calculation] and the results are summarized in Fig.~\ref{FIG:BR} and Table~\ref{Table:BR}.

\begin{table}[tb!]
\begin{center}
\caption{Ir-$L_3$ and $L_2$ edge white-line fitting parameters as determined from pseudo Voigt peak plus sigmoid background function.}
\label{Table:BR}
\resizebox{8.6cm}{!}{
\begin{tabular}{ c  c  c  c  c }
\hline\hline 
   & Center (eV) & Height & Area (eV) & FWHM (eV) \\
\hline
 Peak (pseudo Voigt) &  &  &  & \\
 \hline
  Ir-$L_3$ & 11218.1(1) & 2.24(3) & 20.1(3) & 6.5 \\
  Ir-$L_2$ & 12827.8(1) & 1.52(3) & 13.5(3) & 6.5 \\
\hline  \hline 
 & Center (eV) & Width (eV) & &   \\
\hline
  Step (sigmoid) &  &  &  &  \\
\hline
  Ir-$L_3$ & 11221.2(1) & 6.17(2) &  &  \\
  Ir-$L_2$ & 12836.5(1) & 6.45(2) & &   \\
\hline\hline
\end{tabular}}
\end{center}
\end{table}

The spectra yield a branching ratio of $\sim$ 3.11, which is clearly higher than the statistical BR, but significantly smaller than the BR reported for strongly spin-orbit coupled Ir-oxide systems~\cite{Lagunaprl2010,Clancyprb2012,Kenneyprb2019,Aczelprb2019}. This suggests that the effective SOC strength of Ir in \BNIO~is at most moderate~\cite{Manjilprb2022,Lagunaprl2010,Clancyprb2012,Kenneyprb2019,Aczelprb2019}, and certainly far below the atomic $jj$ coupling limit. Using a BR $\sim 3.11$ and given XANES is a bulk sensitive technique, the average number of holes in the Ir-5$d$ state of \BNIO~is $\left<n_h \right>\approx 5.33$, so the expectation value of the spin-orbit operator $\left<\mathbf{L.S}\right>$ was determined to be 1.34($\hbar^2$), which is smaller than for strongly spin-orbit coupled iridates, but slightly larger than pure Ir metal. The weaker SOC in \BNIO~may not be strong enough to lift the 5$d$-orbital-degeneracy. This is supported by the single peak feature of the nearly symmetric Ir-$L_3$ white line [Fig.\ref{FIG:xanes}(a)] and the corresponding second derivative curve [Fig.~\ref{FIG:xanes}(b)], which should otherwise be asymmetric double peak features corresponding to 2$p_{3/2}$ $\rightarrow$ $J_{\mathrm{eff}} = 5/2$ and $2p_{3/2} \rightarrow J_{\mathrm{eff}} = 3/2$. In addition, the broad linewidth as evident from the values of full-width half maximum (FWHM) (see Table~\ref{Table:BR}) indicates a large bandwidth for the Ir-5$d$ valence orbitals in \BNIO. The combined effects of noncubic crystal distortion, Ir-O covalency, stronger hybridization resulting from direct orbital overlap of the neighboring Ir-5$d$ states, intersite Ir-Ir direct hopping, and large bandwidth suppress the effective SOC strength on Ir in \BNIO. 
{\bf Thus, SOC cannot be the only factor that determines the ground state magnetic and
electronic responses for this system. Together with our band structure calculations, this suggests that the strongly spin-orbit coupled pure $J_{\mathrm{eff}} = \frac{1}{2}$ picture may not be an appropriate for \BNIO\ and that within the face-sharing geometry of the Ir$_3$O$_{12}$ trimer network it may be more appropriate to use a description based on molecular orbital-like states. Such a deviation from the ideal $J_{eff}$ = 1/2 model in our present study is in agreement with the recently reported dimer/trimer based iridium oxide systems consisting of face-sharing Ir-O octahedra \cite{Nagprl2019,Revelli2019,Wangprl2019,Yeprb2018}.}

\subsection{DC and AC magnetic susceptibility and thermodynamic scaling relations}

\begin{figure}[tbh!]
{\includegraphics[width=0.7\linewidth]{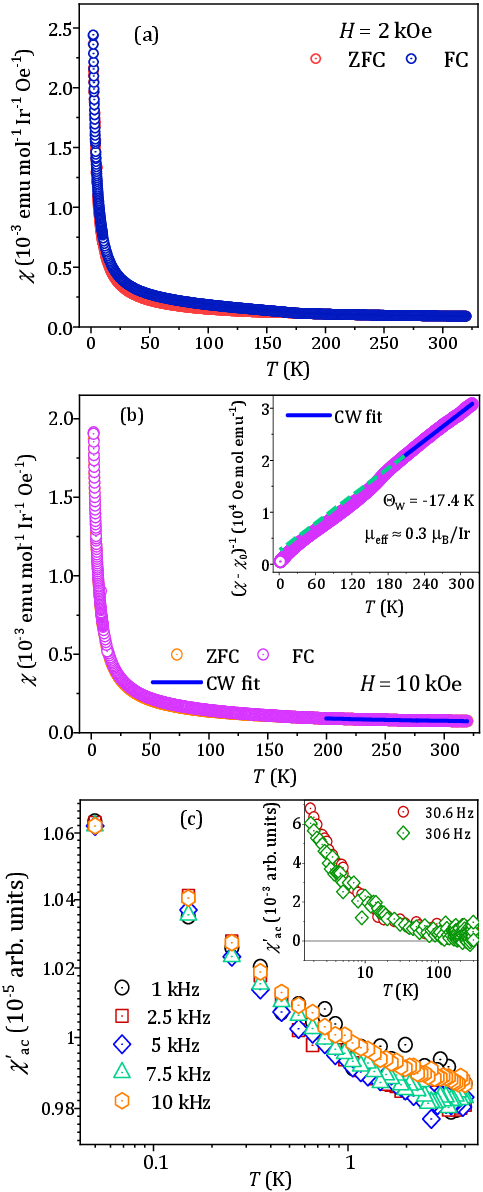}}
\caption{ (a) Zero-field-cooled and field-cooled dc magnetic susceptibility as a function of temperature in applied fields of (a) 2~kOe (a) and (b) 10~kOe along with Curie-Weiss fit (solid blue line) to the FC data. Inset to (b): $T$-dependence of the 10 kOe field-cooled inverse susceptibility, together with a Curie-Weiss fit (blue solid line). The green dashed line indicates a deviation from the CW fit at lowest $T$. (c) Temperature dependence of the real part of the ac magnetic susceptibility on a log-log scale for selected frequencies in a 1~Oe ac excitation field and zero dc bias field. Inset: $T$-dependence of the real part of the ac magnetic susceptibility between 2 and 300~K at two frequencies in a 3~Oe excitation field and a zero dc bias field.}
\label{FIG:Susc1}
\end{figure}

The temperature dependence of the zero-field-cooled (ZFC) and field-cooled (FC) dc magnetic susceptibility of \BNIO~in several applied magnetic fields is shown in Figs.~\ref{FIG:Susc1}(a), \ref{FIG:Susc1}(b) and  Appendix~\ref{Magnetic data}. The featureless paramagnetic-like susceptibility curves, without any ZFC/FC bifurcation in high magnetic fields ($H \geq 5$~kOe) show no sign of magnetic phase transition in \BNIO~down to 2~K. However, for lower applied fields [$H \leq$ 2 kOe as shown in Fig.~\ref{FIG:Susc1}(a)], a noticeable ZFC/FC splitting appears below $\sim$ 175~K. This ZFC/FC divergence at a relatively high temperature in low applied fields could be attributed to the presence of a weakly ferromagnetic Ba-Ir-O (BaIrO$_3$-like) impurity~\cite{bairo3_ref1,bairo3_ref2} that is not detected in either the XRD or EDX measurements. This Ba-Ir-O impurity does not influence the low temperature magnetic response of \BNIO, or the bulk magnetic properties presented below that are determined solely by the desired \BNIO~phase. 

The temperature dependence of ac magnetic susceptibility of \BNIO~was measured at various frequencies down to 0.05~K. The real part of the ac magnetic susceptibility [$\chi^{\prime}_{\mathrm{ac}}(T)$] is illustrated in the main panel of Fig.~\ref{FIG:Susc1}(c). None of these $\chi^{\prime}_{\mathrm{ac}}(T)$ curves display a sharp peak or noticeable frequency dependence that would indicate long-range magnetic order and/or a frozen magnetic ground state. Furthermore, $M(H)$ isotherms, displayed in Appendix~\ref{Magnetic data} for some selected temperatures, show no sign of hysteresis that may indicate a ferromagnetic interaction in this system. In addition, the 2 and 5~K $M^2$ versus $H/M$ Arrott plots shown in Appendix~\ref{Magnetic data}, have a negative intercept on the $M^2$-axis, supporting the absence of any spontaneous magnetization. 

The values of the Curie-Weiss (CW) fitting parameters often depend on the applied magnetic field and the temperature range of the fits~\cite{Nag-JMMM}. Accordingly, CW fits were performed on the field-independent high-$T$ high-field-cooled dc susceptibility~\cite{snioownprb}. Using $\chi = \chi_0 + \frac{C}{(T - \Theta_{\mathrm{W}})}$, where $\chi_0$ is temperature independent susceptibility, and $C$ and $\Theta_{\mathrm{W}}$ are the Curie constant and the Weiss temperature, respectively, Fig.~\ref{FIG:Susc1}(a) and Appendix~\ref{Magnetic data}) {\bf reveal $\chi_0 \sim 4$ to $5 \times 10^{-5}$ emu mol$^{-1}$,} an effective paramagnetic moment, $\mu_{\mathrm{eff}} \sim 0.3 \mu_{\mathrm{B}}$/Ir ion and a significant $\Theta_{\mathrm{W}}$, varying between $-15$ and $-25$~K depending on the applied field and the temperature range of the fit. The effective Ir moment is much smaller than the spin-only $S=1/2$ and the SOC moments of a $J_{\mathrm{eff}} = 1/2$ state ($\approx 1.73~\mu_{\mathrm{B}}$) for a 5$d^5$ Ir$^{4+}$ species. This deviation from the strong SO coupled $J_{\mathrm{eff}} = 1/2$ picture is in line with the lower Ir branching ratio {\bf and the band structure calculations}. The suppression of effective moment could be the result of extended Ir-5$d$ orbitals, direct Ir-5$d$ orbital overlap within the Ir$_3$O$_{12}$ trimer, larger Ir bandwidth, and relatively weak SOC~\cite{Deyprb2012,ba5alir2o9_prb}, suggesting a delocalized nature for the Ir moments in this trimer iridate. The low-temperature saturated moments for the magnetically ordered iridates such as Sr$_2$IrO$_4$, have been found to be $\sim 0.1 \mu_{\mathrm{B}}$/Ir or less while the effective paramagnetic moment is found to be $\sim 0.4 \mu_{\mathrm{B}}$/Ir~\cite{Deyprb2012,Chikaraprb2009}.


\begin{figure*}[tbh!]
\begin{center}
{\includegraphics[width=0.70\linewidth]{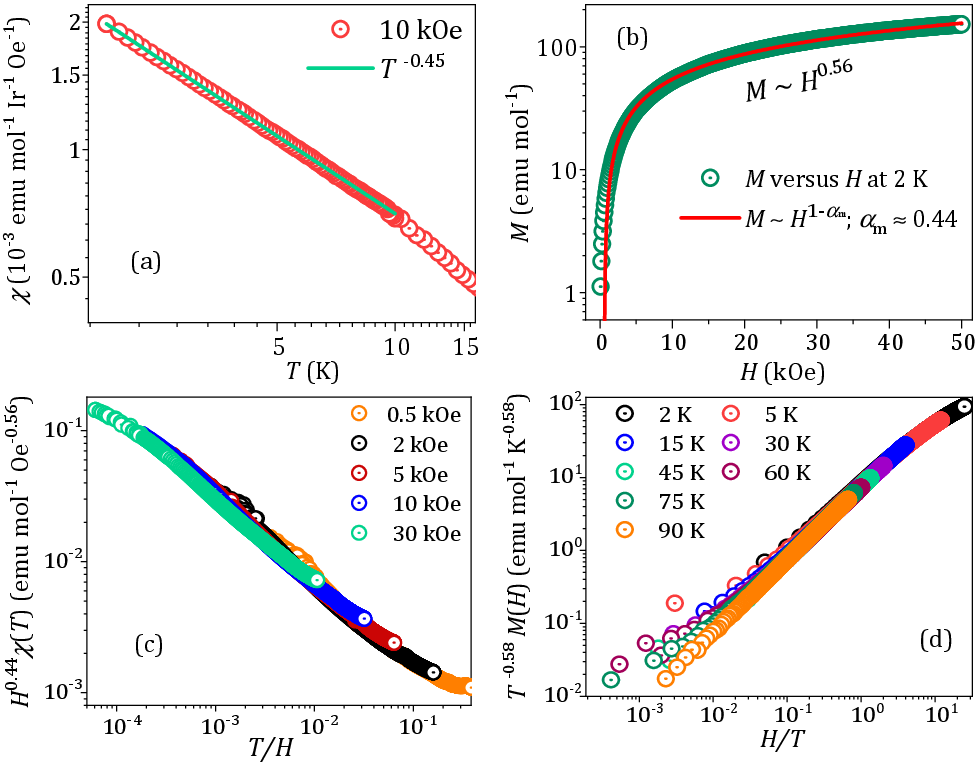}}
\caption{ (a) Temperature dependence of the field-cooled dc magnetic susceptibilityin an applied magnetic field of 10~kOe on a log-log scale, along with a power-law fit (solid green line) in the 2-10~K range. (b) $M - H$ isotherm at 2~K along with a power-law fit, $M(H) \propto H^{1-\alpha_m}$ with $\alpha_m = 0.44$. (c) Scaling of $H^{0.44} \chi_{\mathrm{dc}}(T)$ with $T/H$. (d) Scaled plot of $T^{-0.58}$ $M(H)$ versus $H/T$.}
\label{FIG:Susc2}
\end{center}
\end{figure*}

Figure~\ref{FIG:Susc2}(a) (also see Appendix~\ref{Magnetic data}), shows \BNIO~ exhibits a power-law behavior $\chi(T) \sim T^{-\alpha_{\mathrm{s}}}$ ($\alpha_{\mathrm{s}} = 0.49$ at $H = 500$~Oe and gradually decreases with increasing $H$ to $\alpha_{\mathrm{s}} = 0.39$ for $H = 30$~kOe), instead of following a Curie-tail at low-$T$ ($< 10$~K). Such a sub-Curie dependence suggests finite spin degrees of freedom and rules out any extrinsic paramagnetic impurity as being the origin of the observed low-temperature magnetic response in this material~\cite{cu2iro3prl,h3liir2o6prb}. At 2~K, $M$ versus $H$ also exhibits a power-law dependence, $M(H) \sim H^{1-\alpha_{\mathrm{m}}}$, with $\alpha_{\mathrm{m}} \approx 0.44$ over nearly the entire measured field range [see Fig.~\ref{FIG:Susc2}(b)]. This power-law behavior with $\alpha_{\mathrm{s}} \approx \alpha_{\mathrm{m}}$ originates from a
subset of random singlets induced by a small amount of disorder within either a spin-liquid or a valence-bond-solid ground state~\cite{cu2iro3prl,h3liir2o6prb,Kimchinature2018,Bahramiprl2019,Kimchiprx2018,Knolleprl2019}. Such behavior appears in H$_3$LiIr$_2$O$_6$, LiZn$_2$Mo$_3$O$_8$, ZnCu$_3$(OH)$_6$Cl$_{2}$, Ag$_3$LiIr$_2$O$_6$, and Cu$_2$IrO$_3$ which are QSL materials with disorder~\cite{cu2iro3prl,h3liir2o6prb,Kimchinature2018,Bahramiprl2019,Kenneyprb2019}. To test for a universal scaling relation in \BNIO, $H^{\alpha_{\mathrm{s}}} \chi_{\mathrm{dc}}(T)$ versus $T/H$ and $T^{\alpha_{\mathrm{m}} - 1} M(H)$ versus $H/T$ are plotted in Figs.~\ref{FIG:Susc2}(c) and \ref{FIG:Susc2}(d), respectively. It is evident that the ($H,T$)-dependent $\chi_{\mathrm{dc}}(T)$ data [Fig.~\ref{FIG:Susc2}(c)] overlap over about 4 orders of magnitude in $T/H$ with the same value of $\alpha_{\mathrm{s}} = 0.44$. Similar scaling is seen in the $M(H)$ isotherms (up to $H = 60$~kOe) collected at different temperatures (2 to 90~K) [Fig.~\ref{FIG:Susc2}(d)]. The $T^{\alpha_{\mathrm{m}} - 1} M(H)$ versus $H/T$ curves with $\alpha_{\mathrm{m}} = 0.42$ collapse onto a single scaling curve over nearly five orders of magnitude in $H/T$. The scaling exponents match well between the two distinct thermodynamic quantities $\chi(T)$ and $M(H)$.

\subsection{Muon spin rotation/relaxation}
\begin{figure*}[tbh!]
\centering
{\includegraphics[width=0.70\linewidth]{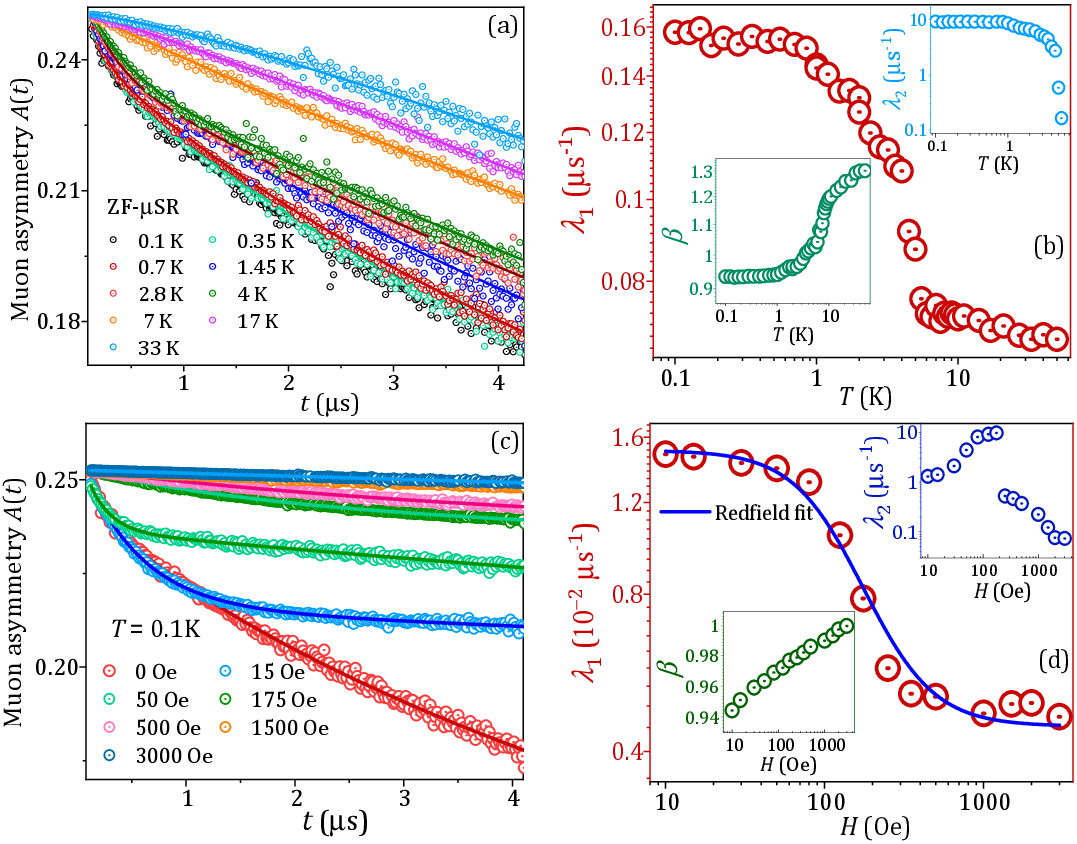}}
\caption{(a) Time evolution of the zero-field muon asymmetry spectra (shaded circles) at selected temperatures, along with the fits (solid lines). (b) Temperature dependence of the slow relaxation rate $\lambda_1$ on a log-log scale, Bottom left inset: Temperature dependence of the stretched exponent $\beta$. Top right inset: Temperature dependence of the fast relaxation rate $\lambda_2$ on a log-log scale. (c) Time evolution of the muon asymmetry at the base temperature of 0.1~K in applied longitudinal fields. (d) Longitudinal field dependence of $\lambda_1$ (main panel), along with the Redfield fitting (blue solid line) on a log-log scale. Bottom left inset: LF dependence of the exponent $\beta$ on a log-linear scale. Top right inset: Log-log plot of the longitudinal field dependence of the fast relaxation rate $\lambda_2$.}
\label{FIG:MUSR}
\end{figure*}

Zero-field and longitudinal-field muon spin rotation/relaxation measurements were performed on \BNIO. Muon spectroscopy is an extremely sensitive microscopic probe of small static local fields (of the order of 0.1~Oe) arising from any weak long-range order or spin freezing. $\mu$SR can also be used to determine the nature of the local spin dynamics and internal magnetic fields of a  magnetically disordered material~\cite{Deyprb2017,snioownprb,Kenneyprb2019}. The time-evolution of the ZF-$\mu$SR asymmetry down to 0.1~K, are shown in Fig.~\ref{FIG:MUSR}(a) for selected temperatures. The spectra gradually change their lineshape from a Gaussian- to a Lorentzian-like relaxation form with decreasing temperature. Down to  0.1~K, the spectra exhibit no signature of static magnetism, with no coherent spontaneous oscillations or 2/3 loss of the initial asymmetry. The ZF-$\mu$SR asymmetry spectra have been fit using
\begin{equation}
A(t) =  A_1\exp(-\lambda_1 t)^\beta + A_2 \exp(-\lambda_2 t) +  A_\mathrm{{bkg}}.
\label{Musrfit}
\end{equation}
The first and second terms in Eq.~\ref{Musrfit} model slow and fast relaxing components, respectively, corresponding to two distinct muon stopping sites, that are likely to be related with the structurally ordered and anti-site disordered Ir-O octahedral coordination environments. The third term is a background contribution originating from the muons stopping in the silver sample holder,


$A_\mathrm{{bkg}}$ was estimated by fitting the ZF-asymmetry at the lowest measured temperature of 0.1~K and then kept fixed at 0.083 throughout the fitting.
The total muon asymmetry was fixed at $\sim 0.256$ while $A_1$ and $A_2$ remain nearly unchanged at $\sim 0.154$ and $\sim 0.019$, respectively, in the temperature window 0.1 to 6~K. Above 6~K, the fast relaxing component disappears ($A_2=0$), increasing the value of $A_1$ to $\sim$ 0.173 for the remaining temperatures (6.5-50~K). At $\leq 6$~K, the relative weighting, $\frac{A_2}{A_1 + A_2} \approx~10.9$\%, is in agreement with the $\sim$10\% Nb/Ir site-exchange obtained from our structural characterization. This suggests that the antisite disordered Ir spins give rise to the fast relaxation, while the structurally ordered correlated Ir moments contribute to the more slowly relaxing signal in the ZF spectra.
Figure~\ref{FIG:MUSR}(b) shows the slower relaxation rate, $\lambda_1$, hardly changes ($\sim0.07-0.075~\mu\mathrm{s}^{-1}$) between 50 and 10~K, consistent with paramagnetic fluctuations of the structurally ordered Ir local moments~\cite{Deyprb2017}. Between $\sim$ 6~and~1~K, $\lambda_1$ increases more rapidly with decreasing temperature, suggesting a slowing down of the site-ordered Ir spin fluctuations through the development of magnetic correlations, as reported in other QSL materials~\cite{Deyprb2017,Liprl2016,Kunduprl2022,Kunduyctoprl2020}. Despite such an obvious slowing down of the spin dynamics, the system does not undergo any static magnetic ordering down to the lowest measured 0.1~K. Instead, below 1 K, $\lambda_1$ saturates (at $\sim 0.15~\mu\mathrm{s}^{-1}$). This temperature-independent plateau [Fig.~\ref{FIG:MUSR}(b)] is representative of the persistent strong quantum spin-fluctuations, within the site-ordered Ir moments in \BNIO~\cite{Deyprb2017,snioownprb,Liprl2016,Kunduprl2022,Kunduyctoprl2020,Clarkprl2013}. The exponent, $\beta$, gradually decreases from $\sim$ 1.3 at 50 K, showing level off at $\beta \approx 0.935$ below 1~K [see left inset in Fig.~\ref{FIG:MUSR}(b)]. The value of $\beta$ close to unity, is much larger than the $\beta = 1/3$ of a canonical spin glass~\cite{Campbellprl}, suggesting nearly homogeneous fluctuating local internal fields around the site-ordered Ir-moments in the trimer. {\bf The observed $\beta > 1$ between 5 and 50~K may originate from trimer singlet fluctuations. Alternatively, the deviation from unity could be due to a distribution of muon sites close to the Nb ions, as discussed in case of the dimer iridate Ba$_3$InIr$_2$O$_9$~\cite{Deyprb2017}.} The right-hand inset of Fig.~\ref{FIG:MUSR}(b) shows that the faster relaxation rate, $\lambda_2$, increases rapidly  below $\sim$ 6~K, reaching a maximum, nearly temperature-independent value ($\sim 9.25~\mu\mathrm{s}^{-1}$)  between 1 and 0.1~K. At these low temperatures, $\lambda_2$ is about two orders of magnitude larger than $\lambda_1$. 

Muon decoupling experiments at 0.1~K in applied longitudinal fields up to 3000~Oe were performed to verify whether the spin correlations are static or dynamic. The results are shown in Fig.~\ref{FIG:MUSR}(c) for selected LFs. The asymmetry curves were fit using Eq.~\ref{Musrfit}, where, $A_{\mathrm{bkg}}$ is fixed at $\sim 0.083$, and $A_1$ and $A_2$ remain nearly field-independent. As shown in the main panel and right-hand inset of Fig.~\ref{FIG:MUSR}(d), $\lambda_1$ gradually decreases with increasing applied LF, while $\lambda_2$, monotonically increases with increasing LF in low fields, possibly due to a rapid slowing down of the structurally disordered Ir-spin-fluctuations at these applied fields, and then exhibits a broad maxima before gradually decreasing. If the muon depolarization arises from any static internal magnetic field of width $\Delta$$H_i$, the zero-field muon relaxation rates, $\lambda_1 \approx 0.155~\mu\mathrm{s}^{-1}$ (slow relaxation) and $\lambda_2 \approx 9.25~\mu\mathrm{s}^{-1}$ (fast relaxation), obtained from the ZF-$\mu$SR data at the base temperature, allow estimates of the size of the static local fields ($\Delta H_i \simeq \lambda_i/\gamma_{\mu}$, where $\gamma_{\mu}$ is the muon gyromagnetic ratio) to be about $\Delta H_1 \approx$ 1.8~Oe and $\Delta H_2 \approx$ 109~Oe. One may expect complete decoupling of the muon spins from the effect of static internal magnetic fields in applied external longitudinal field of (5-10)$\times \Delta H_i$~\cite{Deyprb2017,cu2iro3prl,Kunduprl2022,Kunduyctoprl2020,Middeyqslprb}. It is evident that although there is significant decoupling of the muon spins from the local internal fields even at moderate applied LFs, the $\mu$SR asymmetry spectra undergo perceptible relaxation in the highest applied LF of 3000~Oe [see Figs.~\ref{FIG:MUSR}(c) and \ref{FIG:MUSR}(d)], indicating strongly quantum fluctuating Ir moments down to at least 0.1~K. Further, as demonstrated in the left-hand inset of Fig.~\ref{FIG:MUSR}(d), the exponent, $\beta$, monotonically increases with applied LF and gradually approaches 1 with increasing LF, consistent with other candidate QSL materials. Another interesting observation is that $\lambda_2$ shows a maximum at $H_\textrm{LF} = 175$~Oe and a subsequent drop by two orders of magnitude [see right-hand inset of Fig.~\ref{FIG:MUSR}~(d)]. Clearly, this applied LF of 175~Oe is higher than the estimated static local field ($\Delta H_2 \approx$ 109~Oe), corresponding to the fast relaxation rate of the ZF-muon asymmetry spectrum at the base-temperature.



The LF dependence of the relaxation rates was analysed using the Redfield formula~\cite{Redfield2011}. As shown in the main panel of Fig.~\ref{FIG:MUSR}(d), $\lambda_1$ is well-described with the Redfield formula, while $\lambda_2$ has a more complex structured field variation at lower fields [see right-hand inset to Fig.~\ref{FIG:MUSR}(d)]. As is evident in Fig. \ref{FIG:MUSR}(c), an applied LF $\approx 100$~Oe can suppress the slow relaxation considerably, indicating that the internal fluctuating magnetic fields ($H_{\mu}$) probed by the muons corresponding to this slow relaxation are relatively weak. Estimates using the Redfield formula, $\lambda_1 = \nu \gamma_{\mu}^2 H_{\mu}^2/(\nu^2 + \gamma_{\mu}^2 H_{\mathrm{LF}}^2)$ [Fig.~\ref{FIG:MUSR}(d)] provide the magnitude and frequency of the fluctuating local internal field to be $H_{\mu}\approx 4.8$~Oe and $\nu= 11$~MHz, respectively, which are comparable with other QSL systems~\cite{Liprl2016,Amitavaprb2021,Nagprbdp}. 
{\bf It is to be noted that the generalized Redfield formula fails to describe our $\lambda_1$ vs $H_{LF}$ variation, which probably indicates weakening of correlation among the structurally ordered Ir moments as a result of antisite-disorder.}

The fractional asymmetry-amplitude associated with muon-depolarization via fluctuations does not change with applied LF. This is attributed to the presence of the two distinct muon stopping sites in \BNIO.

The strong geometrical exchange frustration in the edge-shared equilateral triangular network of Ir [see in Figs.~\ref{FIG:Structure}(e) and \ref{FIG:Structure}(f)] facilitates the enhanced quantum spin fluctuations among the Ir moments, so despite a magnetic interaction energy scale $k_{\mathrm{B}}\Theta_{\mathrm{W}}$, no long-range magnetic ordering or a frozen magnetic state is observed down to at least 0.1~K ($\ll k_{\mathrm{B}}\Theta_{\mathrm{W}}$). The lowest measurement temperature of 0.05~K (0.1~K for $\mu$SR) is about several hundred times lower than $\Theta_{\mathrm{W}}$, implying that in spite of some depletion ($\sim 10\%$) in the magnetic equilateral triangular planes due to the Nb-Ir anti-site disorder, the geometrical frustration continues to dominate, stabilizing a dynamic QSL-like fluctuating ground state.

\subsection{Specific heat}
\begin{figure*}[tbh!]
\centering
{\includegraphics[width=0.70\linewidth]{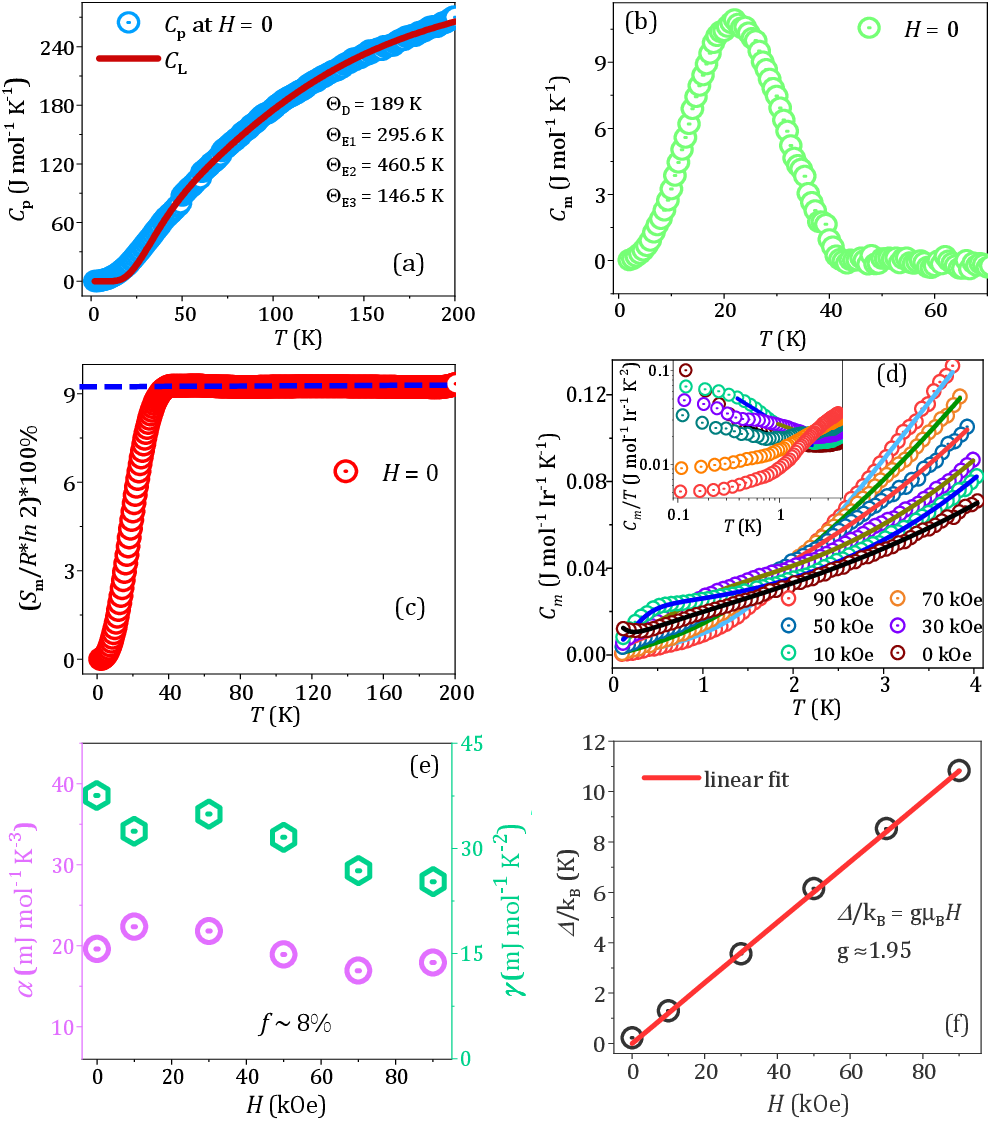}}
\caption{(a) Temperature dependence of the total specific heat $C_{\mathrm{p}}$ between 2 to 200~K in zero magnetic field (open circles) and the lattice heat capacity (red solid line) estimated using a Debye-Einstein model. (b) Temperature dependence of the zero-field magnetic specific heat $C_{\mathrm{m}}$. (c) Magnetic entropy released versus temperature. (d) Temperature dependence of the magnetic specific heat $C_{\mathrm{m}}$ from 0.05 to 4~K in different applied magnetic fields, along with fits (solid coloured lines) using Eq.~\ref{EqHC}. The inset shows $C_{\mathrm{m}}/T$ vs $T$. (e) Variation of $\gamma$ (green open hexagons; right $y$ axis) and $\alpha$ (magenta open circles; left $y$ axis) as a function of applied field. (f) Field dependence of Schottky energy gap obtained from the $C_{\mathrm{m}}(T)$ using Eq.~\ref{EqHC} and a linear fit.}
\label{FIG:HC}
\end{figure*}

The temperature dependence of the specific heat, $C_{\mathrm{p}}(T)$, for \BNIO~in zero magnetic field was measured between 0.05 and 200~K [see Figs.~\ref{FIG:HC}(a) and \ref{FIG:HC}(d)]. There are no anomalies indicating long-range magnetic ordering and/or structural phase transitions down to 50~mK. After subtracting the lattice contribution (see Appendix~\ref{Lattice heat capacity}), the temperature-dependent zero-field magnetic specific heat [Fig.~\ref{FIG:HC}(b)] exhibits a broad peak centered at $\sim 20$~K. This is indicative of frustrated magnetic interactions commonly observed in QSL materials~\cite{okamotoprl2007,Balentsnature,snioownprb,Chengprl2011}. The zero-field magnetic entropy, $S_{\mathrm{m}}$ [Fig.~\ref{FIG:HC}(c)], amounts to only $\sim 9\%$ of the maximum $R\ln 2 \approx 5.76$~J/mol-K$^2$. Even with the reduced effective magnetic moment, $\mu_{\mathrm{eff}} \sim 0.3 \mu_{\mathrm{B}}$/Ir, estimated from the Curie-Weiss analysis, magnetic ordering would give nearly double the magnetic entropy release measured here. This suggests persistent spin fluctuations in the magnetic ground state of this system. 

The very low temperature (0.05 to 4~K) $C_{\mathrm{m}}$ versus $T$ data in both zero and applied magnetic fields is shown in Fig.~\ref{FIG:HC}(d). An upturn is evident in the zero-field $C_{\mathrm{m}}(T)/T$ [see inset to Fig.~\ref{FIG:HC}(d)], which is gradually weakened with increasing field, as observed in many QSL candidates~\cite{Nguyenprm2018,Kellyprx2016,Mustonennature2018,Heltonprl2007,Sheckeltonnature2012,Yamashitanature2008}. $C_{\mathrm{m}}(T)/T$ further exhibits a negative logarithmic temperature dependence in zero field, with a broad hump-like feature developing in an applied $H = 10$~kOe, that shifts to higher temperature with further increase in $H$, possibly indicating a two-level Schottky anomaly that may be due to some free spin centers in this \BNIO~sample. The negative logarithmic temperature dependence of $C_{\mathrm{m}}(T)/T$ in $H \leq 50$~kOe might be an indication of the emergence of quantum critical region~\cite{adrojaqcsl2023,Lohneysen2007}, warranting further field-dependent studies to confirm this. In order to get deeper insight into the nature of low-energy spin-excitations in the dynamic fluctuating ground state of \BNIO, the $C_{\mathrm{m}}(T)$ data between 50~mK and 4~K (both with and without fields) were fit using
\begin{equation}
C_{\mathrm{m}}(T) = \gamma T + \alpha T^2 + f C_{\mathrm{Sch}} + B T (-\ln{T}).
\label{EqHC}    
\end{equation}
\noindent Here $f$ is the fraction of isolated spin centers, $C_{\mathrm{Sch}} = R\left(\frac{\Delta}{k_{\mathrm{B}} T}\right)^2 \frac{\exp\left(\Delta/k_{\mathrm{B}}T\right)}{\left[1 + \exp(\Delta/k_{\mathrm{B}}T)\right]^2}$ is a standard two-level Schottky contribution, and $\gamma$ is the Sommerfeld coefficient typically observed in metals. The results of the fits are summarized in the main panel of Fig.~\ref{FIG:HC}(d). We estimated an $f$ of $\sim$ 8.5\% from the 90~kOe data, then kept $f$ fixed for fits in all other applied fields. This proportion of isolated spin centers is in agreement with the presence of $\sim 8-10\%$ Nb-Ir anti-site-disorder revealed in the structural characterization. The Schottky energy gap $\Delta/k_{\mathrm{B}}$ varies linearly with $H$ [see Fig.~\ref{FIG:HC}(f)], and from the slope of this linear fit the estimated $g$ value of 1.95 matches well with the theoretically calculated Lande-$g$ factor, $g = 2$, for a $S = 1/2$ spin.

$C_{\mathrm{m}}$ exhibits a linear temperature dependence, which together with finite $T$ linear term [$\gamma_{\mathrm{avg}} \approx$~33 mJ mol$^{-1}$K$^{-2}$, shown in the right $y$-axis of Fig.~\ref{FIG:HC}(e)] likely signify low-lying gapless spin excitations from a metal-like spinon Fermi surface of a QSL ground state~\cite{Nagprl2016,Balentsnature,snioownprb,Clarkprl2013,Nagprbdp,Shennature2016,naruo2_nature}. The $\gamma$-value in \BNIO~is comparable to those of the reported in other gapless QSL candidates~\cite{Nagprl2016,snioownprb,Middeyqslprb,Nagprbdp,Mustonennature2018,ba3cusb2o9prl,Yamashitanature2011}. The low-temperature $C_{\mathrm{m}}(T)$ data also show a significant $T^2$-dependency [$C_{\mathrm{m}}(T) \sim \alpha T^2$, with $\alpha \approx 20$~mJmol$^{-1}$K$^{-3}$, shown in the left $y$-axis of Fig.~\ref{FIG:HC}(e)]. Such a $T^2$ dependence of $C_{\mathrm{m}}$ is often observed in the spin-orbit coupled heavier 4$d$ and 5$d$ transition metal based oxides as a fingerprint of a gapless QSL ground state~\cite{Trebstreview2017,Lawlerprl2008,Khuntiaprb2017}. Interestingly, only a combined $\gamma T + \alpha T^2$ term gives a satisfactory fit to each $C_{\mathrm{m}}-T$ data set in the 0.05 to 4~K range. The quadratic $T$-dependence could be the result of Dirac spinon excitations with linear dispersion, similar to the Dirac QSLs Sr$_3$CuSb$_2$O$_9$~\cite{Kunduprl2022} and YbZn$_2$GaO$_5$~\cite{XuarXiv}. A similar $T^2$ dependence of $C_{\mathrm{m}}$ is often seen in frustrated quantum magnets following power-law scaling and data collapse of the thermodynamic quantities~\cite{cu2iro3prl,Kimchinature2018}, which is in consistent with the universal scaling relations here in \BNIO. However, unlike the universal scaling of $\chi(T)$ and $M(H)$, the $C_{\mathrm{m}}[T,H]/T$ data of \BNIO~do not exhibit a power-law-scaling or a data collapse in $T/H$ (see Appendix~\ref{Lattice heat capacity} for the scaling exponent $\alpha = 0.425$). 
\par
{\bf A possible explanation for the lack of scaling behavior in specific heat data could be the presence of additional low-lying excitations in a potential QSL phase through randomness originating from the site exchange. Similar to Kitaev magnet Cu$_2$IrO$_3$~\cite{cu2iro3prl} and very recently reported dimer-based triangular antiferromagnet Ba$_6$Y$_2$Rh$_2$Ti$_2$O$_{17-\delta}$ \cite{Suheon_PRR2024}, in our \BNIO, the Nb/Ir site-disorder induced free spins could modify the low-lying density of magnetic states and consequently the low-energy excitations, resulting in the breakdown of specific heat scaling relation. 
}


\section{Conclusion}
We have investigated the trimer based 12$L$-quadrupolar perovskite iridate \BNIO~through combined structural, electronic, transport, magnetic, thermodynamic, and $\mu$SR measurements. Our combined XRD and Ir-$L_3$ edge EXAFS study indicates the presence of $\sim 8-10\%$ Nb/Ir site-exchange in this material. The Ir $L_{3}$-edge XANES together with Ir-4$f$ core-level XPS confirm that the Ir has a fractional valence of 3.67+, in agreement with the stoichiometrically expected average valence in this material. The valence band x-ray photoemission and temperature-dependent zero-field electrical resistivity data indicate the presence of an energy gap at the Fermi level. Combined dc, ac magnetic susceptibilities, and specific heat measurements reveal that despite having significant antiferromagnetic interactions ($\Theta_W \sim -15$ to $-25$~K) between the finite Ir local moments, the \BNIO~system shows no evidence of magnetic ordering and/or a frozen magnetic state down to at least 50~mK. Detailed zero-field and longitudinal-field $\mu$SR measurements provide evidence for a dynamic quantum spin-liquid ground state in this material.
The universal scaling of two distinct thermodynamic quantities, dc magnetic susceptibility and magnetization, with similar scaling exponents, is consistent with the recent theoretical and experimental studies of spin-liquid systems. The strong geometrical exchange frustration in the edge-shared equilateral Ir-triangular network is expected to induce enhanced quantum-spin-fluctuations in \BNIO, preventing any magnetic ordering down to temperatures much lower than the magnetic exchange energy scale of $k_{\mathrm{B}}\Theta_{\mathrm{W}}$. The magnetic specific heat ($C_{\mathrm{m}}(T)$) data at very low temperature show a linear plus quadratic $T$-dependence, $C_{\mathrm{m}} \sim \gamma T + \alpha T^2$.

We hope that our work will generate additional theoretical and experimental interest in the 3D geometrically frustrated trimer based quadrupole perovskite family of iridates, rhodates and ruthenates as candidate QSL materials.
\section*{Acknowledgements}
A.B and D.T.A would like to thank Professors Liu Hao Tjeng, Andrea Severing, Gheorghe-Lucian  Pascut, Drs Arvid Yogi, Masahiko Isobe, Matthias  Gutmann, Sumanta Chattopadhyay, Gohil Thakur  and Robin Perry for useful discussions. A.B and D.T.A. thank EPSRC UK for the funding (Grant No. EP/W00562X/1). A.B and D.T.A would further like to thank the Royal Society of London for International Exchange funding between the UK and Japan, and Newton Advanced Fellowship funding between UK and China. D.T.A acknowledges the support by the Deutsche Forschungsgemeinschaft (DFG) under the Walter Benjamin Programme, Projektnummer 521584902. The HAXPES and XAS experiments in Japan and Taiwan were supported by the Max Planck-POSTECH-Hsinchu Center for Complex Phase Materials. The authors acknowledge the Materials Characterization Lab (MCL) of ISIS facility, UK for providing the experimental facilities. The authors also thank the ISIS facility for the beam time RB2010177 and Diamond Light Source for beam time on B18 under proposal sp33369. K.Y.C. is supported by the National Research Foundation (NRF) of Korea (Grant No. 2020R1A5A1016518).

\appendix

\section{\label{XRD data data}Powder X-ray diffraction data}
\begin{figure}[tbh!]
\centering
{\includegraphics[width=0.9\linewidth]{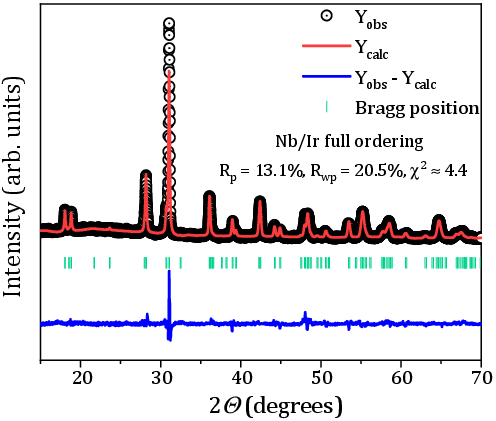}}
\caption{Rietveld refinement on the 300~K XRD pattern considering full Nb-Ir site-ordering.}
\label{FIG:AP1}
\end{figure}

Figure~\ref{FIG:AP1} shows a Rietveld refinement of the XRD pattern of \BNIO~assuming perfect Nb/Ir site ordering. There is poorer agreement between the observed data and the theoretical calculated pattern, along with slightly higher goodness of fit $R$ factors and $\chi^2$ values, than the site disordered model used in Fig.~\ref{FIG:Structure}(a).

\section{\label{Branching ratio (BR)}Ir $L_2$ and $L_3$ white-line analysis, branching ratio (BR) determination, and band structure}

\begin{figure}[tb!]
\centering
{\includegraphics[width=0.75\linewidth]{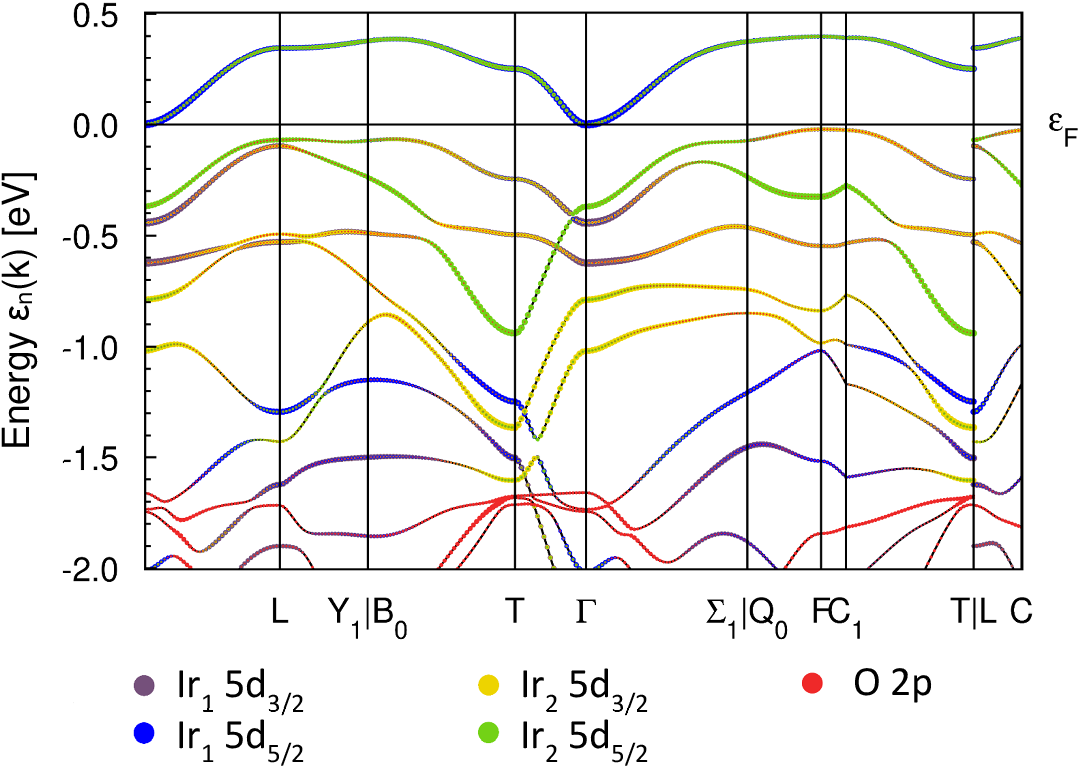}}
\caption{Band structure near the Fermi energy of \BNIO\ calculated with the full potential local orbital (FPLO) basis code. The coloring indicates the band weights projected into the local orbitals. Ir$_1$ and Ir$_2$ are the two distinct Ir-sites within the trimer.}
\label{FIG:AP2}
\end{figure}

{\bf Figure~\ref{FIG:AP2} shows the calculated band structure of \BNIO\ close to Fermi energy. The Ir $t_{2g}$ bands are split into multiple distinct states with highly mixed character, suggesting the formation of mixed orbitals within the Ir$_3$O$_{12}$ trimer. In particular, the strong mixing suggests that this system is far from the ionic $J_{\mathrm{eff}} = 1/2$ picture.} 

To gain a quantitative understanding of the effective SOC in \BNIO~we carefully investigated the Ir-$L_2$ and $L_3$ edge white-line intensities, which here, are dominated by the dipolar transition probabilities: 2$p_{1/2}$ $\rightarrow$ 5$d_{3/2}$ and 2$p_{3/2}$ $\rightarrow$ 5$d_{5/2}$, 5$d_{3/2}$, respectively~\cite{Chojpcm2012,Qiprb1987}. The ratio of these two intensities is the branching ratio, $\mathrm{BR} = \frac{I(L_3)}{I(L_2)}$, which is related to the expectation value of the spin-orbit operator, $\left<\mathbf{L.S}\right>$ via $\mathrm{BR} = \frac{(2+r)}{(1-r)}$, where $r = \frac{\left<\mathbf{L.S}\right>}{n_h}$ and $n_h$ is the average number of holes in the Ir-5$d$ state~\cite{Lagunaprl2010}. The spin-orbit Hamiltonian is given by $H_{\mathrm{SOC}} = \lambda\mathbf{L.S}$, where $\lambda$ is the effective SOC constant. If, in a system, the effective SOC is negligible, the value of BR approaches 2, as $r = \frac{\left<\mathbf{L.S}\right>}{n_h}$, becomes insignificant, and is referred to as the statistical BR. A value of BR considerably higher than 2 implies strong coupling between the local orbital and spin moments~\cite{Manjilprb2022,Lagunaprl2010}.

The experimental BR in \BNIO~was calculated by integrating the resonant cross section at the Ir-$L_2$ and $L_3$ edges after subtracting a step function broadened by the core-hole lifetime to mimic the single-atom absorption process~\cite{Manjilprb2022}. The intensities of the Ir-$L_2$ and $L_3$ absorption edges are normalized for accurate comparison, and the continuum steps at the respective edges are made equal to unity and half of unity, respectively, according to the 1:2 ratio of the occupied 2$p_{1/2}$ and 2$p_{3/2}$ states~\cite{Manjilprb2022}. The white lines have been fit using a pseudo-Voigt function, while a sigmoid function has been used to model the transitions to the continuum at the $L_3$ and $L_2$ absorption edges. The fitting parameters are listed in Table~\ref{Table:BR} and the fit result is shown in Fig.~\ref{FIG:BR}.

\section{\label{Magnetic data}Additional magnetic results}
\begin{figure}[tbh!]
\centering
{\includegraphics[width=0.7\linewidth]{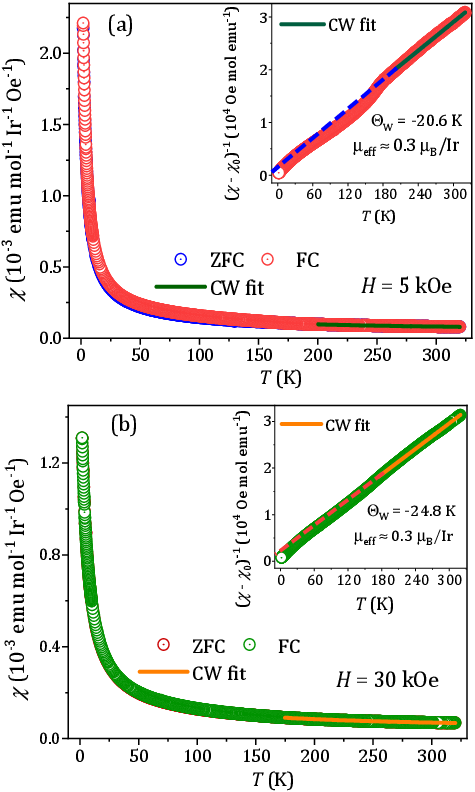}}
\caption{ Zero-field-cooled and field-cooled dc magnetic susceptibility as a function of temperature in applied fields of (a) 5~kOe and (b) 30~kOe along with the Curie-Weiss fits (solid colored lines) on the FC data; Insets show the inverse susceptibility versus temperature along with CW fits, showing deviations from CW behaviour at lower temperature.}
\label{FIG:AP3}
\end{figure}

Figure~\ref{FIG:AP3} shows the temperature dependence of the ZFC and FC dc susceptibility in different applied fields, revealing a paramagnetic behavior. Insets in Fig.~\ref{FIG:AP3} display the Curie-Weiss fits to the inverse susceptibility versus $T$ between 185 and 300~K.

\begin{figure}[tbh!]
\centering
{\includegraphics[width=0.75\linewidth]{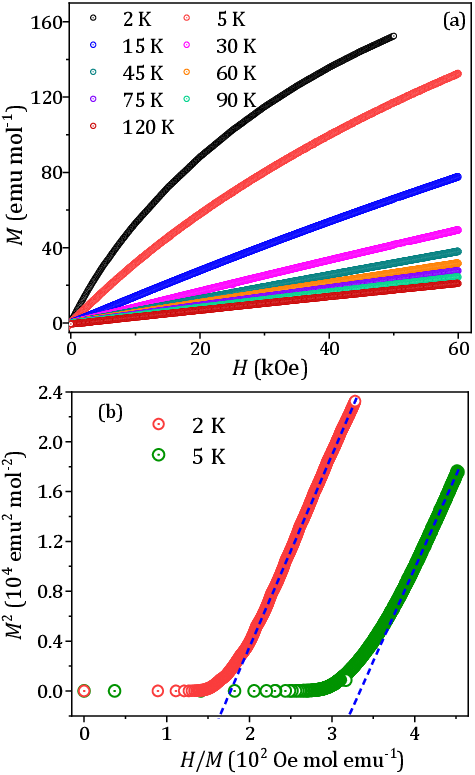}}
\caption{(a) $M$ versus $H$ isotherms at selected temperatures. (b) $M^2$ versus $H/M$ Arrott plots at 2 and 5~K.}
\label{FIG:AP4}
\end{figure}

Figure~\ref{FIG:AP4} shows the first quadrant of the $M$ versus $H$ isotherms at different temperatures and the Arrott plots at 2 and 5 K, revealing the absence of any hysteresis or ferromagnetic interaction in \BNIO.

\begin{figure}[tbh!]
\centering
{\includegraphics[width=0.7\linewidth]{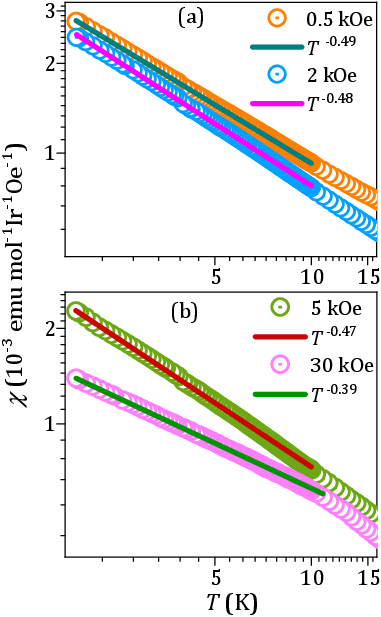}}
\caption{Power-law scaling of the field-cooled dc magnetic susceptibility at 2 to 10~K in applied fields of (a) 0.5 and 2~kOe, and (b) 5 and 30~kOe.}
\label{FIG:AP5}
\end{figure}

Figure~\ref{FIG:AP5} presents the power-law scaling of the 0.5, 2, 5, and 30 kOe field-cooled dc susceptibility data between 2 and 10 K with a scaling exponent varying between $\sim 0.4$ and 0.5.



\section{\label{Lattice heat capacity}Lattice heat capacity extraction and scaling behavior}
As there is no suitable non-magnetic analogue of \BNIO, the lattice contribution to the specific heat, $C_{\mathrm{L}}(T)$ was estimated by fitting the zero-field $C_{\mathrm{p}}$ data in the temperature range 70-200~K using a Debye-Einstein model with a combination of one Debye and three Einstein (1D + 3E) functions [see Fig.~\ref{FIG:HC}(a)]. This yields a Debye temperature $\Theta_{\mathrm{D}} \approx 189$~K and three Einstein temperatures of $\Theta_{\mathrm{E1}} \approx 296$~K, $\Theta_{\mathrm{E2}}\approx 460$~K, and $\Theta_{\mathrm{E3}}\approx 147$~K. $C_{\mathrm{D}}$ and $C_{\mathrm{E}_i} (i = 1-3)$ are the weightings of the acoustic and optical lattice modes, respectively. $C_{\mathrm{D}}:C_{\mathrm{E}_1}:C_{\mathrm{E}_2}:C_{\mathrm{E}_3} \equiv 9.21:4.85:4.87:1.00$, resulting in $C_{\mathrm{D}} + \sum_i C_{\mathrm{E}_i}=19.9$ which is very close to the 20 atoms per formula unit in \BNIO, validating the lattice fitting. This fit is then extrapolated down to the lowest measured temperature and taken as the $C_{\mathrm{L}}$, which was then subtracted from the total $C_{\mathrm{p}}$ to obtain the magnetic specific heat $C_{\mathrm{m}}$. {\bf The small magnetic entropy, just $\sim$ 9\% of R$\ln 2$, could be due in part to an overestimate in the lattice contribution to the specific heat at low temperature. In addition, the relatively high energy scale of the intra-trimer 5$d$ Ir-Ir direct exchange may result in some of the magnetic contribution to the specific heat at higher temperatures being attributed to the lattice.}

\begin{figure}[tbh!]
\centering
{\includegraphics[width=0.7\linewidth]{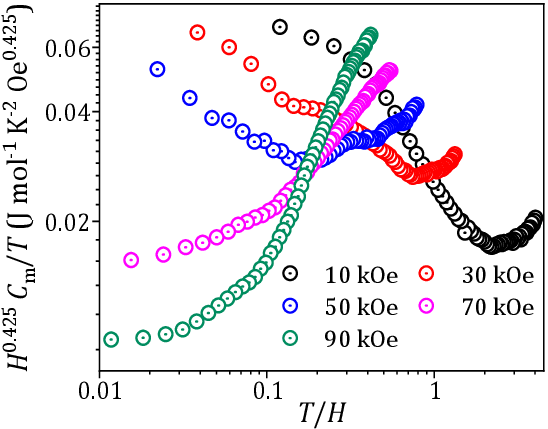}}
\caption{ Deviation from universal scaling behavior in $H^{\alpha_s} C_{\mathrm{m}}/T$ versus $T/H$ with $\alpha_{\mathrm{s}} = 0.425$.}
\label{FIG:AP7}
\end{figure}

Figure~\ref{FIG:AP7} shows the scaling behavior of the magnetic specific heat data measured in various applied magnetic fields. It is clear that the magnetic specific heat data do not follow a universal scaling behavior.

\end{document}